\documentclass[aip,pop,preprint,amsmath,amssymb]{revtex4-1}

\usepackage{graphicx}
\usepackage{amsmath}
\renewcommand{\vec}[1]{\boldsymbol{#1}}

\newcommand{\BM}{Box--Muller }
\newcommand{\MJ}{Maxwell--J\"{u}ttner }

\def\apj{{\itshape Astrophys. J.} }
\def\jgr{{\itshape J. Geophys. Res.} }
\def\pop{{\itshape Phys. Plasmas} }
\def\ppcf{{\itshape Plasma Phys. Control. Fusion} }

\begin{document}

\title{Loading a relativistic kappa distribution in particle simulations}

\author{Seiji Zenitani}
\affiliation{Research Center for Urban Safety and Security, Kobe University, 1-1 Rokkodai-cho, Nada-ku, Kobe 657-8501, Japan.}
\email{zenitani@port.kobe-u.ac.jp}
\author{Shin'ya Nakano}
\affiliation{The Institute of Statistical Mathematics, 10-3 Midori-cho, Tachikawa, Tokyo 190-8562, Japan}

\date{Submitted to Physics of Plasmas}

\begin{abstract}
A procedure for loading particle velocities from a relativistic kappa distribution in particle-in-cell (PIC) and Monte Carlo simulations is presented.
It is based on the rejection method and the beta prime distribution.
The rejection part extends earlier method for the \MJ distribution,
and then the acceptance rate reaches $\gtrsim 95\%$.
Utilizing the generalized beta prime distributions,
we successfully reproduce the relativistic kappa distribution, including the power-law tail.
The derivation of the procedure, mathematical preparations, comparison with other procedures, and numerical tests are presented.
\end{abstract}

\maketitle

\section{Introduction}

The kappa distribution is one of the most fundamental velocity distributions
in kinetic studies in space and solar-wind plasmas.\citep{book}
Since it was introduced in 1960s,\citep{vas68,olbert68}
the kappa distribution has been widely used to study
plasmas with suprathermal populations,
because it seamlessly contains both the thermal Maxwellian component and
the nonthermal power-law component in the high-energy part.
It is also noteworthy that
the kappa distribution is connected with Tsallis statistical mechanics,
as it maximizes a non-extensive entropy.\citep{livadiotis09,tsallis88,book}
For the theory and applications of the kappa distribution,
the readers may refer to a recent monograph\citep{book} and references therein. 
Kinetic plasma processes in a kappa-distributed space plasma
have been investigated
by using particle-in-cell (PIC) simulations.\citep{lu10,koen12,park13,abdul15,abdul21}
To simulate a kappa-distributed plasma,
one often has to initialize particle velocities according to the kappa distribution,
however, numerical procedures to load kappa distributions may not be well-documented.
Among them, \citet{abdul15} recognized that the kappa distribution is
equivalent to the multivariate $t$-distribution,
which can be generated from a normal distribution and
a chi-squared distribution.\citep{kroese11}

To deal with energetic electrons of $\gtrsim 0.5$ MeV and energetic ions of $\gtrsim 1$ GeV,
special relativity needs to be considered.
Velocity distributions need to be modified accordingly.
A relativistic Maxwell distribution,
often referred to as a \MJ distribution,\citep{jut11}
has been used in Monte Carlo simulations
as well as in PIC simulations in high-energy astrophysics. 
It is not straightforward to generate a \MJ distribution,
because the relativistic Lorentz factor $\gamma=[1-(v/c)^2]^{1/2}$
makes the problem difficult.
To load \MJ distributions, several rejection-based algorithms have been proposed.\citep{sobol76,poz83,canfield87,sk13,swisdak13,zeni15b}
For example, \citet{sobol76} has proposed a simple rejection method,
based on the gamma distribution.
\citet{canfield87} have utilized four gamma distributions.
Their method achieves good acceptance rate of $\gtrsim 70\%$ for nearly arbitrary initial conditions.
Another option is the inverse transform method,
which refers to a numerical table of the cumulative distribution function.
In such a case, one needs to carefully adjust the size of the table,
because the code often becomes inefficient.

A relativistic kappa distribution will be useful
in modeling energetic processes in laser, space, and astrophysical plasmas.
Its mathematical form including the normalization constant was
provided by \citet{xiao06} and by \citet{lht22}.
Tsallis-type statistics in relativistic collisionless plasmas
has gained attention very recently.\citep{zhdankin22}
Meanwhile, to the best of our knowledge, no one has presented
a numerical procedure to load a relativistic kappa distribution
in particle simulations.
One can similarly consider the inverse transform method,
but a much larger table will be required,
because the kappa distribution has a power-law tail that extends nearly infinitely.
We desire a reliable algorithm that is free from tables.

In this article, we propose a numerical algorithm to load
a relativistic kappa distribution in PIC and Monte Carlo simulations.
The rest of this manuscript is organized as follows.
As starting points,
Sections \ref{sec:maxwell} and \ref{sec:kappa} discuss
algorithms to generate Maxwell and kappa distributions.
Section \ref{sec:MJ} presents our extention of \citet{canfield87}'s algorithm
to generate a \MJ distribution.
Section \ref{sec:rkappa} introduces
a new procedure to generate a relativistic kappa distribution,
based on beta prime distributions.
Section \ref{sec:tests} presents numerical tests of the proposed methods for relativistic distributions.
The efficiency of the rejection method are evaluated.
Section \ref{sec:discussion} contains discussions and summary.

\section{Maxwell distribution}
\label{sec:maxwell}

A Maxwell distribution is a multivariate normal distribution.
\begin{align}
\label{eq:maxwell}
f_{\rm M}(\vec{v})d^3{v}
&= N_M \Big(\frac{1}{\pi v_M^2}\Big)^{\frac{3}{2}} \exp \Big( -\frac{ \vec{v}^2 }{v_M^2} \Big) d^3{v}
\end{align}
Here, $N_M$ is the number density for a Maxwellian plasma, $v_M$ is the most probable velocity, and the other symbols have their standard meanings. In this case, plasma temperature is given by $T_M = (1/2) m v_M^2$. Throughout the paper, we focus on isotropic distributions.

The three components of the Maxwellian can be obtained by
\begin{align}
\label{eq:multi_n}
{v}_x = \sigma n_1,
~~
{v}_y = \sigma n_2,
~~
{v}_z = \sigma n_3,
\end{align}
where $\sigma^2=(1/2)v^2_{\rm M}$ is the variance and
$n_1, n_2,$ and $n_3$ are the normal random variates.
The normal variates can be generated by the \BM method\citep{bm58} or other methods.\citep{devroye86,kroese11,yotsuji10}

We examine the Maxwellian from another angle for discussions in later sections.
We move to spherical coordinates
via $d^3v = 4 \pi v^2 dv$.
We use a normalized parameter
\begin{align}
x &\equiv \frac{v^2}{v^2_M},
\end{align}
and then we obtain
\begin{align}
f_{\rm M}(x) dx
= \frac{2 N_M}{\sqrt{\pi} } x^{1/2} e ^{-x} dx
= N_M {\rm Ga}\left(x; \frac{3}{2}, 1\right) dx
\end{align}
where $\Gamma(x)$ is the gamma function and
\begin{align}
{\rm Ga}(x; k, \lambda) =
\frac{ x^{k-1} e^{ -x / \lambda } }{ \Gamma(k) \lambda^{k}}
\end{align}
is the gamma distribution with a shape parameter $k$ and a scale parameter $\lambda$.
There are several procedures to generate the gamma distributions.\citep{devroye86,mt00,kroese11,yotsuji10}
We present some of them in Appendix \ref{sec:gamma}.
From a gamma-distributed random variate $X_{{\rm Ga}(\alpha,\beta)}$,
we recover the velocity
\begin{align}
v = v_M \sqrt{ X_{{\rm Ga}(3/2,1)} }
= \sigma\sqrt{ X_{{\rm Ga}(3/2,2)} }
.
\label{eq:maxwell_gam23a}
\end{align}
We can also rewrite this with $T_M$, 
\begin{align}
v
= \sqrt{ \frac{2T_M}{m} X_{{\rm Ga}(3/2,1)} }
= \sqrt{ X_{{\rm Ga}(3/2,2T_M/m )} }
.
\label{eq:maxwell_gam23b}
\end{align}
Finally, we obtain the $\vec{v}$ vector,
by randomly scattering $v$ onto the spherical surface,
using two uniform random variates $X_1,X_2 \sim U(0,1)$.
\begin{eqnarray}
\label{eq:spherical_scattering}
\left\{ \begin{array}{lll}
v_x &=& v ~ ( 2 X_1 - 1 ) \\
v_y &=& 2 v \sqrt{ X_1 (1-X_1) } \cos(2\pi X_2) \\
v_z &=& 2 v \sqrt{ X_1 (1-X_1) } \sin(2\pi X_2)
\end{array} \right.
\end{eqnarray}

\section{Kappa distribution}
\label{sec:kappa}

A kappa distribution is defined by
\begin{align}
\label{eq:kappa}
f_{\rm \kappa}(\vec{v})  d^3v
= \frac{N_{\kappa}}{(\pi\kappa\theta^2)^{3/2}} \frac{\Gamma(\kappa+1)}{\Gamma(\kappa-1/2)}  \Big( 1 + \frac{ \vec{v}^2 }{\kappa \theta^2} \Big)^{-(\kappa+1)} d^3v
\end{align}
where $\theta$ is the most probable speed and $\kappa$ is the kappa parameter.
The $\kappa$ parameter controls a power-law index in the high-energy part.
It ranges $\kappa > 3/2$, so that the effective plasma temperature $T = [\kappa / (2\kappa-3)] m\theta^2$ remains finite.
Other properties of the kappa distribution are discussed in \citet{book} and references therein.
Assuming isotropy, we obtain
\begin{align}
\label{eq:kappa_v}
f_{\rm \kappa}(v) dv
&= N_{\kappa} \frac{4}{\pi^{1/2}(\kappa\theta^2)^{3/2}} \frac{\Gamma(\kappa+1)}{\Gamma(\kappa-1/2)}  \Big( 1 + \frac{ {v}^2 }{\kappa \theta^2} \Big)^{-(\kappa+1)} v^2 dv \\
&= N_{\kappa} {\rm B'}\left( v;\frac{3}{2},\frac{\nu}{2},2,({\kappa}\theta^2)^{1/2} \right) dv
\label{eq:kappa_betap}
.
\end{align}
Here, $\nu=2\kappa-1$ and
${\rm B'}$ is the generalized beta prime distribution
with a shape parameter $p$ and a scale parameter $q$,
\begin{align}
\label{eq:gbetap}
{\rm B'}(x;\alpha,\beta,p,q)
&=
\frac{p}{q B(\alpha,\beta)}
\left(\frac{x}{q}\right)^{\alpha p - 1}
\left( 1 + \left(\frac{x}{q}\right)^p \right)^{-(\alpha+\beta)}
\nonumber \\
&=
\frac{p\Gamma(\alpha+\beta)}{q^{\alpha p} \Gamma(\alpha)\Gamma(\beta)}
\left( 1 + \left(\frac{x}{q}\right)^p \right)^{-(\alpha+\beta)}x^{\alpha p - 1}
.
\end{align}
$B(\alpha,\beta)$ is the beta function.
If we redefine $x' \equiv (x/q)^p$, with help from $(dx'/x')=p(dx/x)$,
then $x'$ follows the (standard) beta prime distribution,
\begin{align}
\label{eq:betap}
{\rm B'}(x';\alpha,\beta)
&=
{\rm B'}(x';\alpha,\beta,1,1)
=
\frac{\Gamma(\alpha+\beta)}{\Gamma(\alpha)\Gamma(\beta)}
\left( 1 + {x'} \right)^{-(\alpha+\beta)}(x')^{\alpha - 1}
.
\end{align}
A random variate according to the beta prime distribution,
$X_{{\rm B'}(\alpha,\beta)}$, is generated by
\begin{align}
X_{{\rm B'}(\alpha,\beta)}
=
\frac{
X_{{\rm Ga}(\alpha,\delta)}
}{
X_{{\rm Ga}(\beta,\delta)}
}
\label{eq:delta}
\end{align}
where $X_{{\rm Ga}(\alpha,\delta)}$ is a random variate
that follows the gamma distribution ${\rm Ga}(\alpha,\delta)$.\citep{forbes10}
Here Eq.~\eqref{eq:delta} is independent of choice of $\delta$.
Since the relation (Eq.~\eqref{eq:delta}) is not well known,
we provide a brief proof in Appendix \ref{sec:beta}.
Then, considering the scaling factors, we obtain
\begin{align}
\label{eq:gbetap_random}
X_{{\rm B'}(\alpha,\beta,p,q)}
=
q\left( X_{{\rm B'}(\alpha,\beta)} \right)^{1/p}
=
q
\left(
\frac{
X_{{\rm Ga}(\alpha,\delta)}
}{
X_{{\rm Ga}(\beta,\delta)}
}
\right)^{1/p}
\end{align}
Setting $\delta=2$,
the kappa-distributed velocity $v$ is given by
\begin{align}
v =
\left(
\kappa \theta^2
\frac{
X_{{\rm Ga}(3/2,2)}
}{
X_{{\rm Ga}(\nu/2,2)}
}
\right)^{1/2}
\label{eq:v_recovery}
\end{align}
Then we generate two gamma distributions.
We note that
the shape parameter $\nu/2$ in the denominator
can be an integer, half-integer, or floating-point number,
as long as it satisfies $\nu/2=\kappa-1/2>1$.
Gamma generators for arbitrary $k>1$ can be found
in Appendix \ref{sec:gamma} and references therein \citep{devroye86,mt00,kroese11,yotsuji10}. 
After this, we spherically scatter $\vec{v}$ to the $v_x, v_y$, and $v_z$ components
by Eq.~\eqref{eq:spherical_scattering}.
The entire steps to generate the kappa distribution are presented in Algorithm 1-1 in Table \ref{table:kappa}.

In Section \ref{sec:maxwell}, we have already seen that
the gamma distribution with $k=3/2$ and the spherical scattering provide the Maxwell distribution (Eqs.~\eqref{eq:multi_n}, \eqref{eq:maxwell_gam23a}, and \eqref{eq:spherical_scattering}).
From Eqs.~\eqref{eq:v_recovery} and \eqref{eq:spherical_scattering},
we similarly obtain the three components of the kappa distribution:
\begin{align}
\label{eq:kappa1}
{v}_x = \sqrt{\kappa\theta^2}~\frac{n_1}{\sqrt{\chi^2_\nu}},
~~
{v}_y = \sqrt{\kappa\theta^2}~\frac{n_2}{\sqrt{\chi^2_\nu}},
~~
{v}_z = \sqrt{\kappa\theta^2}~\frac{n_3}{\sqrt{\chi^2_\nu}},
\end{align}
where $n_1, n_2,$ and $n_3$ are the normal random variates and
${\chi^2_\nu} = X_{{\rm Ga}(\nu/2,2)}$ is a random variate
according to the chi-squared distribution with $\nu$ degrees of freedom.
The procedure is presented in Algorithm 1-2 in Table \ref{table:kappa}. 
Mathematically, the kappa distribution is identical to a multivariate $t$-distribution in three dimensions.\citep{abdul15}
It is known that a multivariate $t$-distribution can be generated from a multivariate normal distribution divided by a chi-squared distribution with $\nu$ degrees of freedom.\citep{kroese11}
Algorithm 1-2 is equivalent to this well-known procedure.

\begin{table*}
\begin{center}
\caption{Algorithms to generate a kappa distribution.
$\kappa > 3/2$ is required.
See also Appendix \ref{sec:gamma} to generate gamma-distributed variates.
\label{table:kappa}}
\begin{minipage}{0.45\textwidth}
\begin{tabular}{l}
\hline
{\bf Algorithm 1-1}\\
\hline
generate $X_1, X_2 \sim U(0, 1)$ \\
generate $X_3 \sim {\rm Ga}(3/2, 2)$ \\
generate $\chi^2_\nu \sim {\rm Ga}(\kappa - 1/2, 2)$ \\
$v \leftarrow \sqrt{ X_3 \dfrac{ \kappa \theta^2 }{ \chi^2_\nu }}$\\
$v_x \leftarrow v ~ ( 2 X_1 - 1 )$ \\
$v_y \leftarrow 2 v \sqrt{ X_1 (1-X_1) } \cos(2\pi X_2)$ \\
$v_z \leftarrow 2 v \sqrt{ X_1 (1-X_1) } \sin(2\pi X_2)$ \\
\hline
\end{tabular}
\end{minipage}
\begin{minipage}{0.45\textwidth}
\begin{tabular}{l}
\hline
{\bf Algorithm 1-2}\\
\hline
generate $n_1, n_2, n_3 \sim \mathcal{N}(0, 1)$ \\
generate $\chi^2_\nu \sim {\rm Ga}(\kappa - 1/2, 2 )$ \\
$r \leftarrow \sqrt{ \dfrac{\kappa\theta^2}{ \chi^2_\nu } }$ \\
$v_x \leftarrow r n_1$ \\
$v_y \leftarrow r n_2$ \\
$v_z \leftarrow r n_3$ \\
\hline
\\
\end{tabular}
\end{minipage}
\end{center}
\end{table*}

\begin{figure}[htbp]
\centering
\includegraphics[width={\columnwidth}]{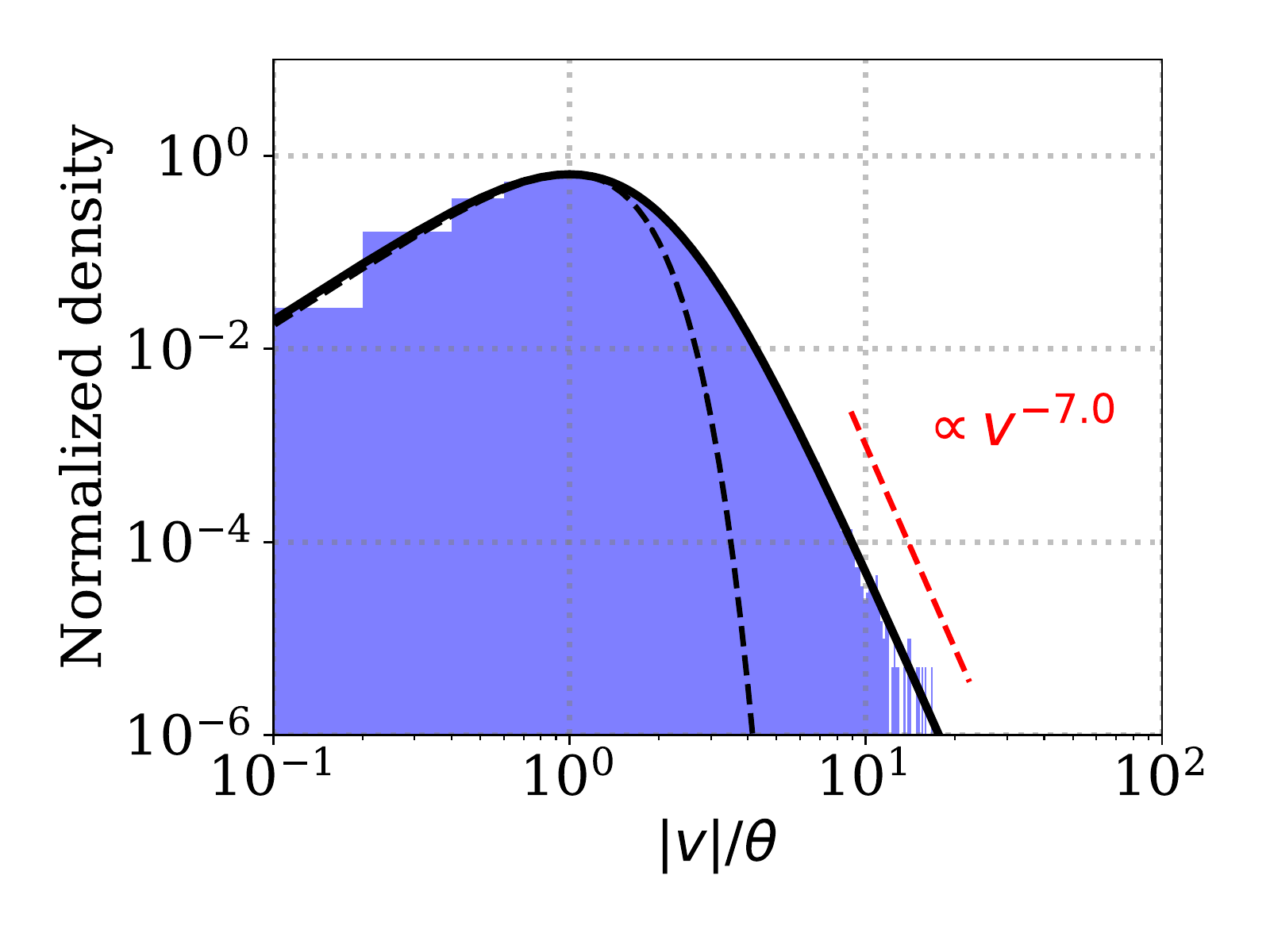}
\caption{
Velocity distribution of kappa-distributed particles with $\kappa=3.5$ (the blue histogram)
and its analytic solution (the black line).
The black dashed line indicates the adjusted Maxwellian.
\label{fig:kappa_v}}
\end{figure}

\begin{figure}[htbp]
\centering
\includegraphics[width={\columnwidth}]{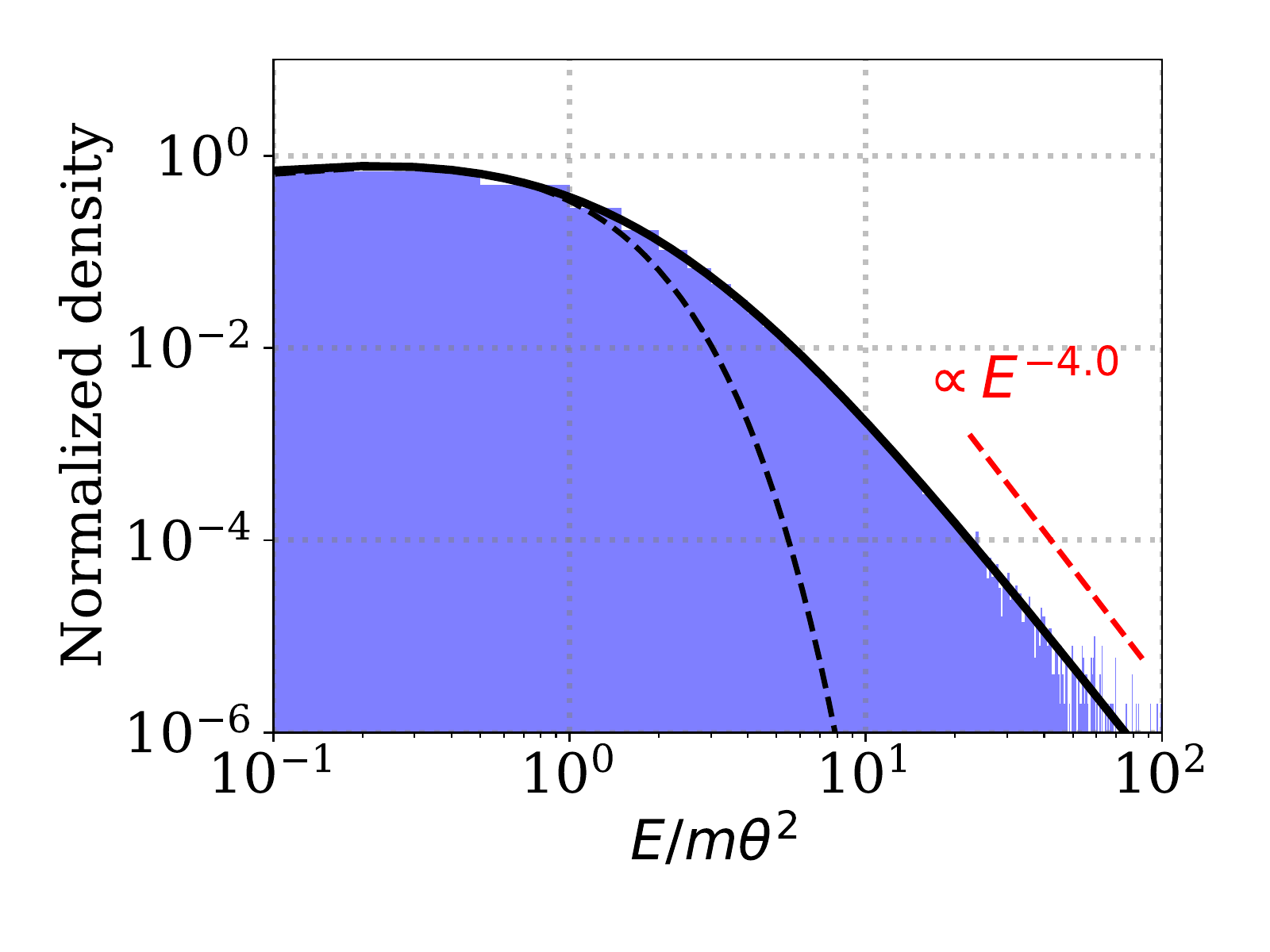}
\caption{
Energy distribution of kappa-distributed particles with $\kappa=3.5$ (the blue histogram)
and its analytic solution (the black line).
The black dashed line indicates the adjusted Maxwellian.
\label{fig:kappa_E}}
\end{figure}

We have checked the algorithm by using $10^6$ random particles.
Generating kappa-distributed particles with $\kappa=3.5$,
we have made histograms of $f(v)$ in Fig.~\ref{fig:kappa_v} and
$f(E)$ in Fig.~\ref{fig:kappa_E}.
Particle count numbers (the blue histograms) and analytic curves (the black lines) are in excellent agreement.
From Fig.~\ref{fig:kappa_v}, it is evident that $v=\theta$ is the most probable velocity.
From Eq.~\eqref{eq:kappa_v}, we can see that a power-law tail in the high-energy part scales like $f(v) \propto v^{-2\kappa}$.
After small mathematics, we also see $f(E) \propto E^{-(\kappa+1/2)}$.
As indicated by the red dashed lines, properties of the power-law tail is excellently reproduced.
To better visualize the high-energy tail,
we overplot a Maxwell distribution
with $v_{M'}=\theta$ and $N_{M'} = 0.78 N_{\kappa}$
by the black dashed line.
These parameters are carefully set such that
the Maxwellian is inscribed to the kappa distribution
at the most probable speed of $v = \theta$,
as described in Appendix \ref{sec:in_kappa}.
We pay attention to the difference between the tail (the solid line) and the inscribed Maxwellian (the dashed line) for $v \ge \theta$.
We define it the nonthermal part.
For $\kappa \lesssim 3.5$, more than a half of the kinetic energy is carried by the nonthermal particles.\citep{oka13}

\section{Relativistic Maxwell distribution}
\label{sec:MJ}

We describe an algorithm to load
a relativistic Maxwell distribution, also known as a \MJ distribution.
Since the procedure extends an earlier work by \citet{canfield87},
we call it the modified Canfield method.

The phase-space density of a \MJ distribution is given by
\begin{equation}
\label{eq:MJ}
f_{\rm MJ}(\vec{p})d^3{p}
= \frac{N_M}{4\pi m^2c T_M K_2 (mc^2/T_M)} \exp \Big( -\frac{ \gamma mc^2 }{T_M} \Big) d^3{p}
\end{equation}
where $\gamma=[1-(v/c)^2]^{-1/2}$ is the Lorentz factor, $\vec{p}=m\gamma \vec{v}$ is the momentum, and $K_n(x)$ is the modified Bessel function of the second kind. 
Hereafter we set $m=c=1$ and write $t \equiv {T_M}/{mc^2}$ for brevity.
In the spherical coordinates, we obtain
\begin{align}
\label{eq:exp_u2}
f_{\rm MJ}(p) dp
=
\frac{N_M}{t K_2 (1/t)}
\exp \Big(-\frac{\sqrt{1+p^2} }{t} \Big) p^2 dp
\end{align}
Using a kinetic energy parameter $x$
\begin{align}
\label{eq:x}
x &\equiv \frac{{E}}{mc^2} = {\gamma-1}
,
\end{align}
we rewrite Eq.~\eqref{eq:exp_u2} into
\begin{align}
f_{\rm MJ}(x) dx
&=
\frac{e^{-1/t}}{t K_2 (1/t)}
e^{-x/t} ~ ( 1+x )\sqrt{ ( x+1 )^2 - 1} ~ dx
\nonumber \\
&=
\frac{e^{-1/t}}{t K_2 (1/t)}
e^{-x/t} ~ ( \sqrt{2}x^{1/2} + ax + b\sqrt{2}x^{3/2} + x^2 )~
\frac{(1+x)\sqrt{x+2}}{\sqrt{2} + a x^{1/2} + b\sqrt{2}x + x^{3/2}}
~ dx 
\label{eq:MJ_detail}
\end{align}
where we have introduced two hyperparameters: $a$ ($0 < a \le 1$) and $b$ ($0 < b \le 1$).
Unless stated otherwise, we employ $a=0.56$ and $b=0.35$ throughout this paper.

Following \citet{canfield87},
we define weight functions $w_i(t)$, 
their sum $S(t)$, normalized weights $\pi_i(t)$, and
a rejection function $R(x)$,
\begin{align}
\label{eq:weight}
w_3(t) &= \sqrt{\pi}
,~~
w_4(t) = a\sqrt{2t}
,~~
w_5(t) = \frac{3b\sqrt{\pi}t}{2}
,~~
w_6(t) = {(2t)^{3/2}},\\
S(t) &= \sum_{i=3}^6 w_i(t) =
\sqrt{\pi} + a\sqrt{2t} + \frac{3b\sqrt{\pi}t}{2} + {(2t)^{3/2}} \\
\pi_i(t) &= \frac{ w_i(t) }{ S(t) }
~~~(i=3,4,5,6),
\label{eq:MJ_prob}
~~~\sum_{i=3}^6 \pi_i(t)=1,
\\
R(x; a, b) &\equiv \frac{(1+x)\sqrt{x+2}}{\sqrt{2} + a {x}^{1/2} + b\sqrt{2}x + x^{3/2}}
\label{eq:rej}
\end{align}
This rejection function ranges $0.957 \lesssim R(x) < 1$.
It is asymptotic to $1$
in the $x \rightarrow 0$ and $x \rightarrow \infty$ limits.

We also remind the readers of the gamma distribution with shape $k/2$ and scale $\lambda$,
\begin{align}
\label{eq:chi}
Ga(x; k/2,\lambda) = \frac{1}{\lambda^{k/2}\Gamma(k/2)} x^{k/2 - 1} e^{-x/\lambda}
\end{align}
In particular, we take advantage of the following four distributions,
\begin{align}
\label{eq:chi3456}
Ga\Big(x;\frac{3}{2},t\Big) &= \frac{2}{\sqrt{\pi}t^{3/2}} x^{1/2} e^{-x/t}
,~~~
Ga(x;2,t)  = \frac{1}{t^2} x e^{-x/t}
,\nonumber\\
Ga\Big(x;\frac{5}{2},t\Big) &= \frac{4}{3\sqrt{\pi}t^{5/2}} x^{3/2} e^{-x/t}
,~~
Ga(x;3,t) = \frac{1}{2t^3} x^{2} e^{-x/t}
,
\end{align}
Using Eqs.~\eqref{eq:weight}--\eqref{eq:rej} and \eqref{eq:chi3456},
Eq.~\eqref{eq:MJ_detail} yields
\begin{align}
f_{\rm MJ}(x) dx
&=
\frac{\sqrt{t}e^{-1/t} S(t)}{\sqrt{2}  K_2 (1/t)}
\left( \sum_{i=3}^6 \pi_i(t) ~Ga\Big(x;\frac{i}{2},t\Big) \right)
R(x)
~ dx
\label{eq:canfield87}
\end{align}
We also define an auxiliary distribution $F(x)$ without $R(x)$
\begin{align}
F_{\rm MJ}(x) dx
&=
\frac{\sqrt{t}e^{-1/t} S(t)}{\sqrt{2} K_2 (1/t)}
\left( \sum_{i=3}^6 \pi_i(t)
~Ga \Big(x;\frac{i}{2},t\Big) \right)
~ dx
\label{eq:envelope}
\end{align}

Since $\sum_{i=3}^6 \pi_i(t)=1$,
the bracketed term in Eq.~\eqref{eq:envelope} stands for a superposition of four gamma distributions at the probability of $\pi_i(t)$. 
Using a uniform random variate $X_1 \sim U(0,1)$ and the probability $\pi_i(t)$,
one can choose one gamma distribution from the four.
A random variate for the gamma distribution ${\rm Ga}(x; i/2,t)$
can be easily generated, for example, by the procedure in Appendix \ref{sec:gamma}.
Superimposing four distributions,
we obtain the auxiliary distribution $F_{\rm MJ}(x)$.
Then we apply a rejection method. 
Using another uniform random $X_2 \sim U(0,1)$,
we accept $x$ when the following condition is met:
\begin{align}
X_2 < R(x)
\label{eq:rej_x2}
\end{align}
If this condition is not met, we reject the particle, and
then we regenerate the random variate.
We further use the squeeze technique to evaluate Eq.~\eqref{eq:rej_x2}.
As will be shown, the rejection function is greater than 0.95.
Considering this, we rewrite the condition,
\begin{align}
X_2 < 0.95 ~~{\bf or}~~ X_2 < R(x)
\label{eq:squeeze}
\end{align}
Then, the code immediately returns true without calling the rejection function in 95\% of the cases,
and it calls the rejection function only in the remaining 5\% cases.
This usually makes the code faster.

After the rejection, we obtain $x$ according to the distribution of Eq.~\eqref{eq:canfield87}.
We translate $x$ into $p$
\begin{align}
\label{eq:x2p}
p = \sqrt{x(x+2)}
\end{align}
and then we further convert it to $\vec{p}$ via Eq.~\eqref{eq:spherical_scattering}.
The obtained $\vec{p}$ follows the \MJ distribution in 3-D momentum space (Eq.~\eqref{eq:MJ}).
Algorithm 2 in Table~\ref{table:MJ} displays
the entire procedure in the pseudo code.
The normalization constant in Eq.~\eqref{eq:MJ} does not appear in the procedure.

\begin{table}
\begin{center}
\caption{The modified Canfield method to generate a \MJ distribution.
See also Appendix \ref{sec:gamma}.
\label{table:MJ}}
\begin{tabular}{l}
\\
\hline
{\bf Algorithm 2: the modified Canfield method}\\
\hline
$a \leftarrow 0.56,~~b \leftarrow 0.35,~~R_0 \leftarrow 0.95$\\
compute $\pi_{3},\pi_{4},\pi_{5}$ for given $t$ using Eqs.~\eqref{eq:weight}--\eqref{eq:MJ_prob}\\
{\bf repeat}\\
~~~~generate $X_1, X_2 \sim U(0, 1)$ \\
~~~~{\bf if}~~~~~~$X_1 < \pi_{3}$ ~~~~~~~~~~~~~~~{\bf then} $i \leftarrow 3$\\
~~~~{\bf elseif} $X_1 < \pi_{3}+\pi_{4}$ ~~~~~~~~{\bf then} $i \leftarrow 4$\\
~~~~{\bf elseif} $X_1 < \pi_{3}+\pi_{4}+\pi_{5}$ ~{\bf then} $i \leftarrow 5$\\
~~~~{\bf else} $i \leftarrow 6$ \\
~~~~{\bf endif}\\
~~~~generate $x \sim {\rm Ga}( i/2,t )$\\
{\bf until} $X_2 < R_0$ {\bf or} $X_2 < R(x;a,b)$
\\
generate $X_3, X_4 \sim U(0, 1)$ \\
$p \leftarrow \sqrt{ x \left( x+2 \right) }$ \\
$p_x \leftarrow p ~ ( 2 X_3 - 1 )$ \\
$p_y \leftarrow 2 p \sqrt{ X_3 (1-X_3) } \cos(2\pi X_4)$ \\
$p_z \leftarrow 2 p \sqrt{ X_3 (1-X_3) } \sin(2\pi X_4)$ \\
\hline
\end{tabular}
\end{center}
\end{table}

\begin{figure}[htbp]
\centering
\includegraphics[width={\columnwidth}]{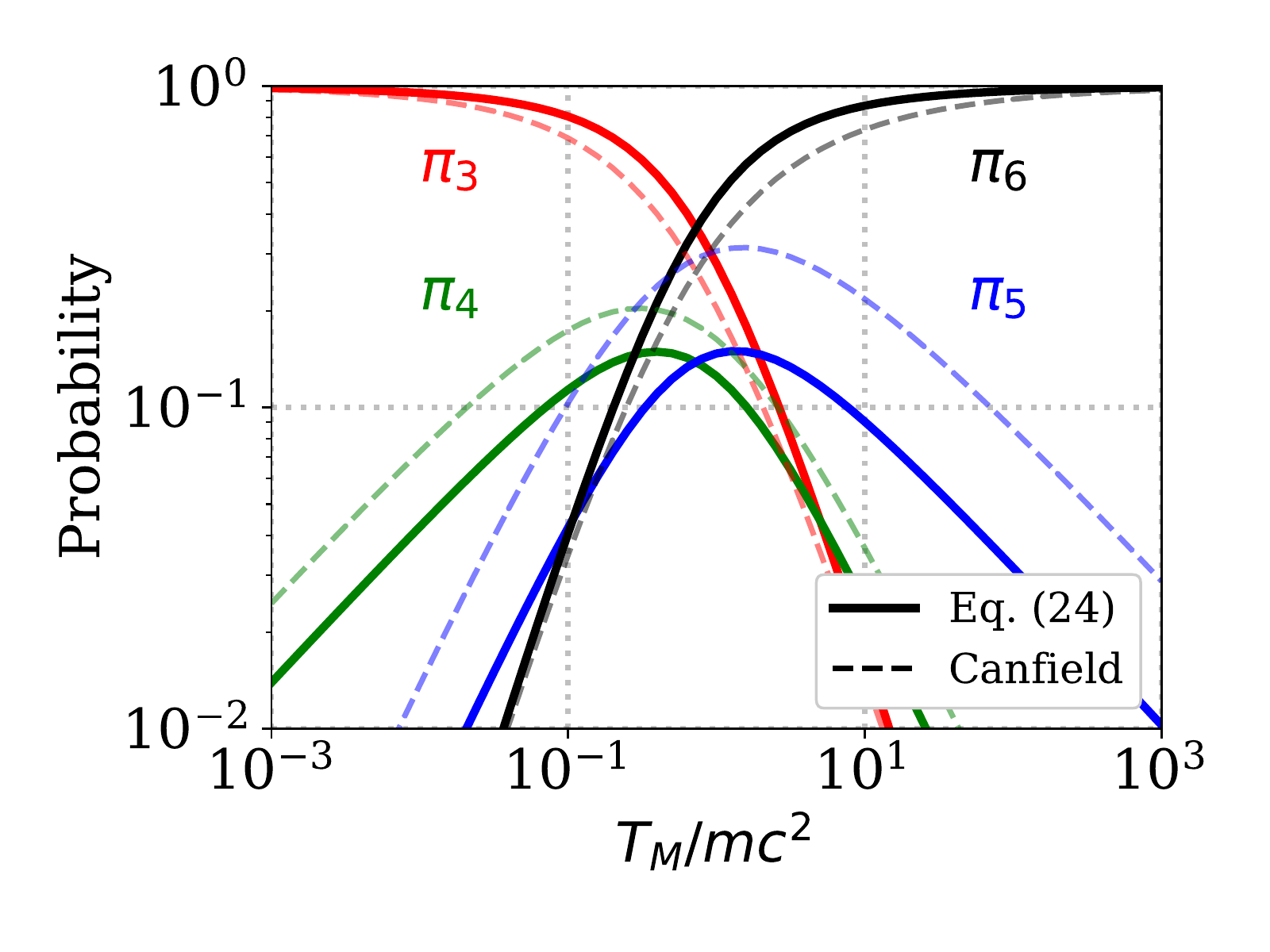}
\caption{
Probability densities $\pi_3, \pi_4, \pi_5, \pi_6$ as a function of the relativistic temperature $t=T_M/mc^2$.
The thick line indicates the probabilities by the new method (Eq.~\eqref{eq:MJ_prob}; $a=0.56, b=0.35$).
The thin dashed line indicates the probabilities in the original Canfield method ($a=1, b=1$).
\label{fig:MJ_prob}}
\end{figure}

\begin{figure}[htbp]
\centering
\includegraphics[width={\columnwidth}]{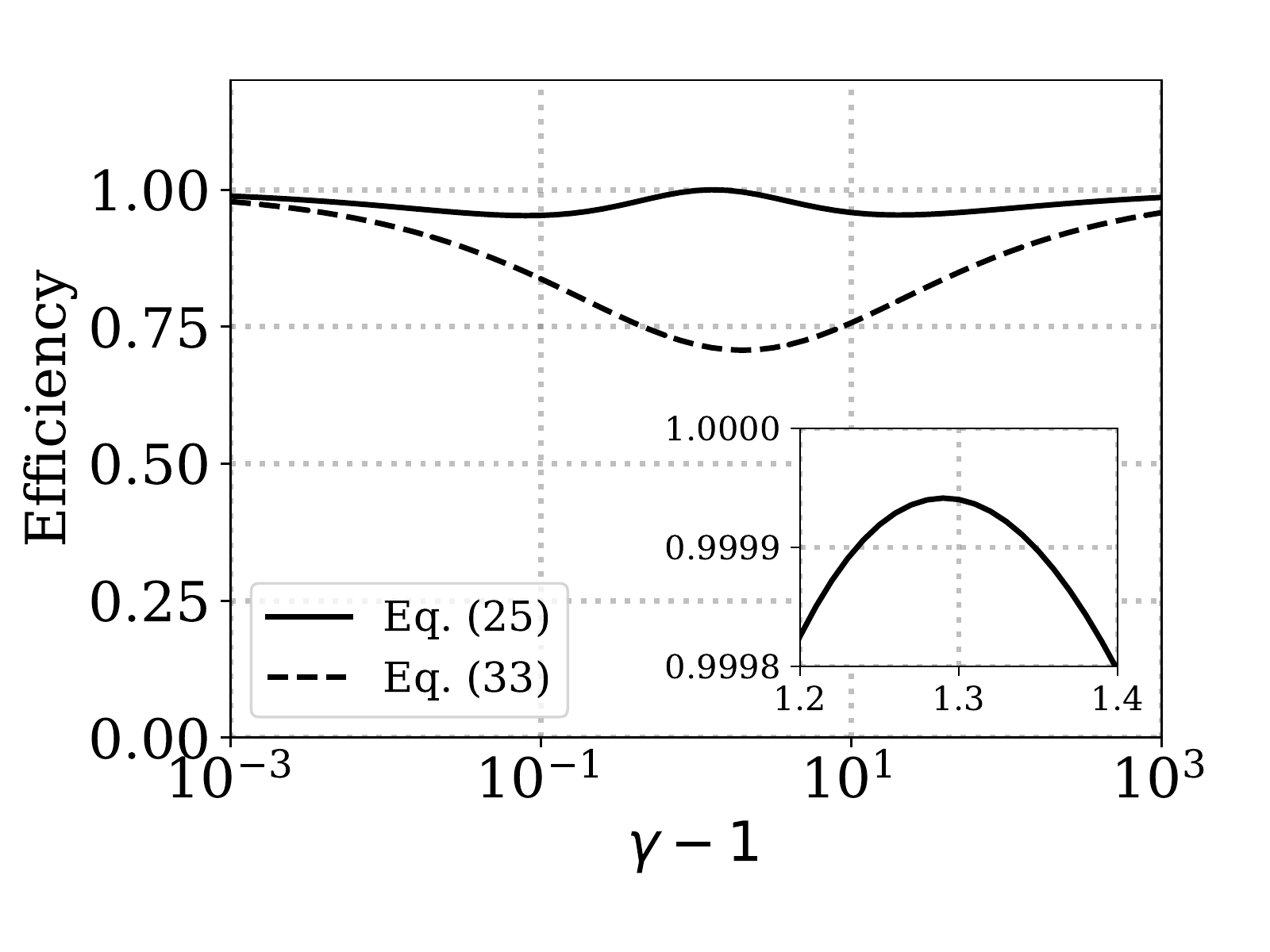}
\caption{
Comparison of the rejection functions.
The new function (Eq.~\eqref{eq:rej}; $a=0.56, b=0.35$) and the original function (Eq.~\eqref{eq:cf_rej}). The inlet zooms in on the peak of the new function.
\label{fig:rej}}
\end{figure}

Behavior of this algorithm depends on
the plasma temperature,
because the probability $\pi_i(t)$ (Eq.~\eqref{eq:MJ_prob}) is a function of $t=T_M/mc^2$.
Fig.~\ref{fig:MJ_prob} shows
the $t$-dependence of the probability densities.
The thick solid lines indicate $\pi_i$ in color.
The four components gradually change their percentages.
In the nonrelativistic temperature limit of $t \ll 1$,
the $\pi_3$ component is dominant.
In such a case, from Eqs.~\eqref{eq:chi3456} and \eqref{eq:x2p},
we estimate $v \simeq (2x)^{1/2} \approx \sqrt{ 2 X_{{\rm Ga}(3/2,t)} } = \sqrt{ X_{{\rm Ga}(3/2,2t)} }$.
This recovers the nonrelativistic Maxwell distribution,
as discussed in Eq.~\eqref{eq:maxwell_gam23b} in Section \ref{sec:maxwell}.
As the temperature increases,
the $\pi_3$ component decreases.
Instead, the $\pi_4$ and $\pi_5$ components appear successively.
The three components are replaced by the $\pi_6$ component, which is dominant in the $t \gg 1$ limit.
In this limit, the distribution (Eq.~\eqref{eq:exp_u2}) approaches $f_{\rm MJ}(p)dp \approx \exp( - p / t ) p^2 dp$, and $p \approx x$ (Eq.~\eqref{eq:x2p}). This is consistent with the $Ga(x;3,t)$ term in Eq.~\eqref{eq:chi3456}.
Compared with the $\pi_3$ and $\pi_6$ components,
the $\pi_4$ and $\pi_5$ components are localized in the medium temperature range.

The original method by \citet{canfield87} corresponds to the $a=b=1$ case without the squeeze method.
Quantities in Eqs.~\eqref{eq:weight}--\eqref{eq:rej} need to be recalculated accordingly.
The probabilities $\pi_i$ for $a=b=1$ are shown by the thin dashed lines in Fig.~\ref{fig:MJ_prob}.
From the original (dashed lines) to the new (solid lines) methods,
the $\pi_4$ and $\pi_5$ components decrease,
because we rescale them by factors of $a$ and $b$.
Instead, the $\pi_3$ and $\pi_6$ components increase relatively.
The rejection function (Eq.~\eqref{eq:rej}) for the original Canfield method is reduced to be
\begin{align}
\label{eq:cf_rej}
R_{\rm CF}(x) = R(x;1,1) = \frac{\sqrt{x+2}}{\sqrt{x}+\sqrt{2}}
.
\end{align}
Fig.~\ref{fig:rej} compares the rejection functions, Eq.~\eqref{eq:cf_rej} ($a=1, b=1$) and Eq.~\eqref{eq:rej} ($a=0.56, b=0.35$).
The original function ranges from $1/\sqrt{2} \le R(x) < 1$,
and it has a minimum at $x=2$.
The Canfield method relatively inefficient around $1\lesssim x \lesssim 10$. 
To overcome this, we have reduced the $\pi_4$ and $\pi_5$ components that are localized in the medium range by the factors $a, b \in [0,1]$.
We have found our choice of $(a,b)=(0.56, 0.35)$ by a grid-search method, as described in Appendix \ref{sec:grid}.
Using these parameters,
the new rejection function becomes surprisingly efficient, as presented in Fig.~\ref{fig:rej}.
Its two minimum values are almost equal,
$R(x) \approx 0.9571$ at $x \approx 0.053$ and
$\approx 0.9578$ at $x \approx 9.85$.
The local maximum at $x \approx 1.29$ is very close to, but smaller than the unity, as shown in the inlet.
One can verify our parameters $(a,b)=(0.56, 0.35)$ from different angle.
Considering $R(x;a,b)\approx 1$, we expect both $R(1;a,b) = 2\sqrt{3} / (a + \sqrt{2}b + 1 + \sqrt{2}) \approx 1$ and $R(2;a,b) = 3 \sqrt{2} / (a + {2}b + 3 ) \approx 1$.
By solving them, we obtain a good estimate $(a,b) \approx (0.58, 0.33)$.

\begin{figure}[htbp]
\centering
\includegraphics[width={\columnwidth}]{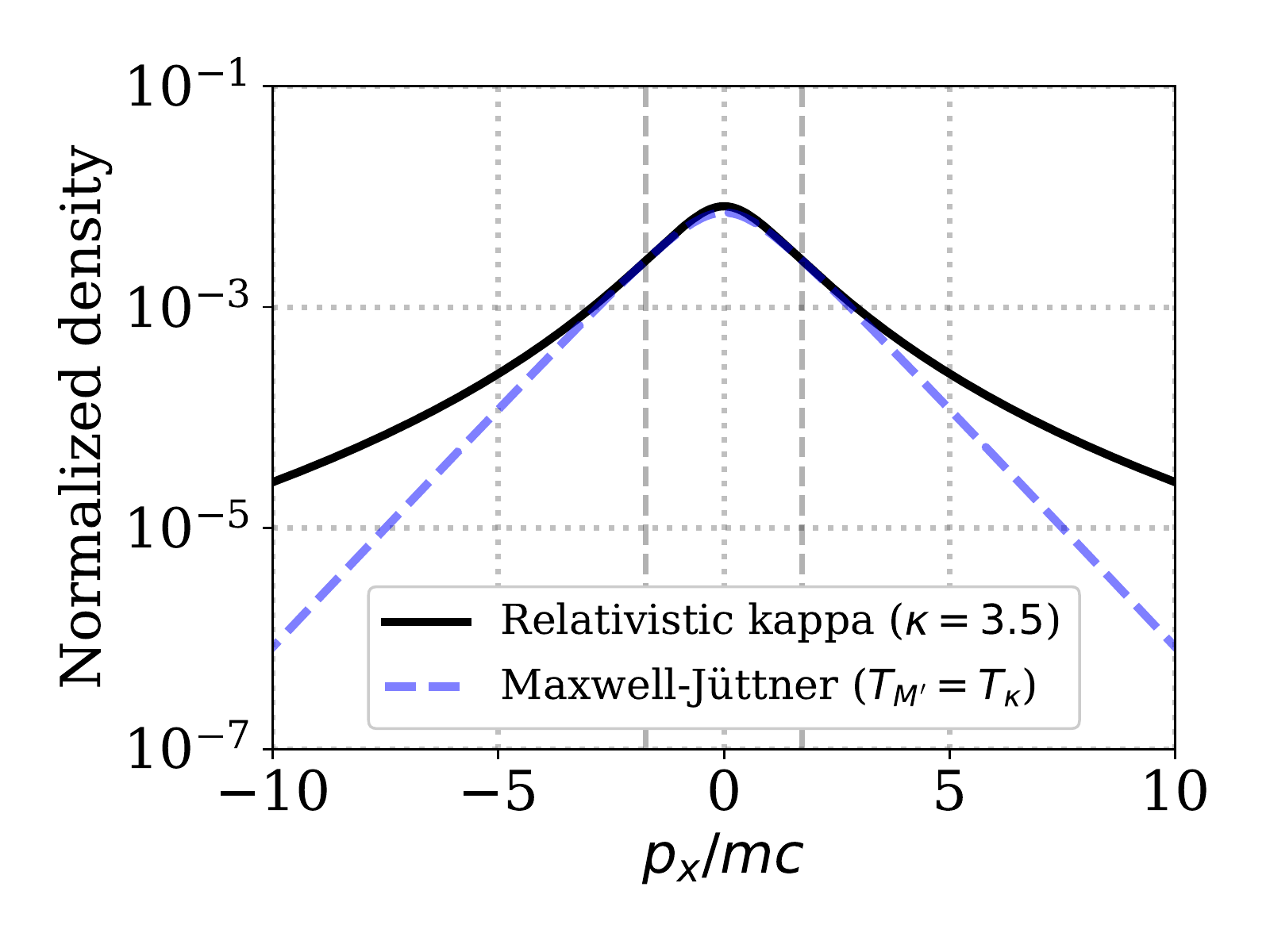}
\caption{
Relativistic kappa distribution with $T_{\kappa}/mc^2=1$ and $\kappa=3.5$ (black line).
The blue dashed line shows an adjusted \MJ distribution
with $T_{M'}=T_{\kappa}$. 
\label{fig:rkappa_core}}
\end{figure}

\section{Relativistic kappa distribution}
\label{sec:rkappa}

A relativistic kappa distribution is defined by\citep{lht22}
\begin{align}
\label{eq:rkappa}
f_{\rm RK}(\vec{p}) d^3p
&\equiv 
A \left( 1 + \frac{ (\gamma-1)mc^2 }{\kappa T_{\kappa}} \right)^{-(\kappa+1)}  d^3p\\
A(\kappa,T_{\kappa})
& =
\frac{
N_{\kappa}
\Gamma\left( \kappa +  \frac{1}{2} \right)
}{
{(2\pi m \kappa T_{\kappa})^{3/2}}
(\kappa+1)
~\Gamma( \kappa - 2 )
~{}_2F_1\Big( -\frac{3}{2}, \frac{5}{2}; \kappa+\frac{1}{2}; 1-\frac{\kappa T_{\kappa}}{2mc^2}\Big)
}
\label{eq:rkappa_constant}
\end{align}
where $A$ is the normalization constant,
$N_{\kappa}$ is the plasma density,
$T_{\kappa}$ is a characteristic temperature,
and ${}_2 F_1$ is the hypergeometric function.
The $\kappa$ parameter should satisfy $\kappa > 3$,
so that the energy density remains finite.
We set $m=c=1$ and write $t \equiv T_{\kappa} / mc^2$ for brevity.
The black solid line in Fig.~\ref{fig:rkappa_core} shows
a 1-D cut of a relativistic kappa distribution with $\kappa=3.5$ and $t=1$. 
The blue dashed line indicates a \MJ distribution
with $T_{M'} = T_{\kappa}$ and $N_{M'} = 0.40 N_{\kappa}$.
This distribution is inscribed to the relativistic kappa distribution
at the energy of ${E}=(\gamma-1)mc^2=T_{\kappa}$
(at $p_x=\pm \sqrt{3}$; the gray vertical lines),
as described in Appendix \ref{sec:in_rkappa}.
Compared with the \MJ distribution,
the relativistic kappa distribution has broader high-energy tails.

Moving to the spherical coordinates, we obtain
\begin{align}
\label{eq:rkappa_u2}
f_{\rm RK}({p})d{p}
& =
A(\kappa,t) \left( 1 + \frac{ \gamma-1 }{\kappa t} \right)^{-(\kappa+1)}
4 \pi p^2 ~ dp
\end{align}
We use the normalized kinetic energy $x$,
\begin{align}
\label{eq:x2}
x &\equiv \frac{E}{mc^2} = {\gamma-1}
\end{align}
and then Eq.~\eqref{eq:rkappa_u2} yields
\begin{align}
f_{\rm RK}(x) dx
&= 4\pi A(\kappa,t)
\left( 1 + \frac{ x }{\kappa t} \right)^{-(\kappa+1)}
 ~ \Big( 1+{x} \Big)\sqrt{ \Big( 1+{x} \Big)^2 - 1} ~ dx \\
&= 4\pi A(\kappa,t)
\left( 1 + \frac{ x }{\kappa t} \right)^{-(\kappa+1)}
\Big( \sqrt{2}{x}^{1/2} + a{x} + b\sqrt{2}{x}^{3/2}+{x}^2 \Big)
\frac{(1+x)\sqrt{x+2}}{\sqrt{2} + a x^{1/2} + b\sqrt{2}x + x^{3/2}}
~ dx
\label{eq:rkappa_detail}
\end{align}
We set the hyperparameters to $a=0.56$ and $b=0.35$.

We define weight functions $w_i(\kappa,t)$, 
their sum $S(\kappa,t)$, normalized weights $\pi_i(\kappa,t)$, and
a rejection function $R(x)$,
\begin{align}
w_3(\kappa,t)
&= 
\frac{\sqrt{2\pi}}{2}
{\Gamma\Big(\kappa-\frac{1}{2}\Big)}
,~~~~~~~~~~~~~
w_4(\kappa,t)
= 
a \sqrt{\kappa t}~
\Gamma(\kappa-1)
,
\nonumber \\
w_5(\kappa,t)
&= 
\frac{3b\sqrt{2\pi}}{4}
{(\kappa t)~
\Gamma\Big(\kappa-\frac{3}{2}\Big)}
,~~~~
w_6(\kappa,t)
= 
2{(\kappa t)^{3/2}~
\Gamma(\kappa-2)}
, \label{eq:rkappa_weight}
\\
S(\kappa,t) &\equiv
\sum_{i=3}^6 w_i(\kappa,t)
, \label{eq:rkappa_sum}
\\
\pi_i(\kappa,t) &= \frac{w_i(\kappa,t)}{S(\kappa,t)}
~~~(i=3,4,5,6),
~~~\sum_{i=3}^6 \pi_i(\kappa,t)=1,
\label{eq:rkappa_prob} \\
R(x; a, b) &\equiv \frac{(1+x)\sqrt{x+2}}{\sqrt{2} + a {x}^{1/2} + b\sqrt{2}x + x^{3/2}}
. \label{eq:rkappa_rej}
\end{align}
In this section, we use the generalized beta prime distributions,
\begin{align}
{\rm B'}\Big(x;\frac{3}{2},\kappa-\frac{1}{2},1,\kappa t\Big)
&=
\frac{\Gamma(\kappa+1)}{(\kappa t)^{3/2}~
\Gamma(3/2)\Gamma(\kappa-1/2)}
\left( 1 + \frac{x}{\kappa t} \right)^{-(\kappa+1)}x^{1/2}
, \label{eq:rakappa_beta3} \\
{\rm B'}\Big(x;2,\kappa-1,1,\kappa t\Big)
&=
\frac{\Gamma(\kappa+1)}{(\kappa t)^2~
\Gamma(2)\Gamma(\kappa-1)}
\left( 1 + \frac{x}{\kappa t} \right)^{-(\kappa+1)}x
, \label{eq:rakappa_beta4} \\
{\rm B'}\Big(x;\frac{5}{2},\kappa-\frac{3}{2},1,\kappa t\Big)
&=
\frac{\Gamma(\kappa+1)}{(\kappa t)^{5/2}~
\Gamma(5/2)\Gamma(\kappa-3/2)}
\left( 1 + \frac{x}{\kappa t} \right)^{-(\kappa+1)}x^{3/2}
, \label{eq:rakappa_beta5} \\
{\rm B'}\Big(x;3,\kappa-2,1,\kappa t\Big)
&=
\frac{\Gamma(\kappa+1)}{(\kappa t)^{3}~
\Gamma(3)\Gamma(\kappa-2)}
\left( 1 + \frac{x}{\kappa t} \right)^{-(\kappa+1)}x^2
. \label{eq:rakappa_beta6}
\end{align}
Using Eqs.~\eqref{eq:rkappa_weight}--\eqref{eq:rakappa_beta6},
Eq.~\eqref{eq:rkappa_detail} yields,
\begin{align}
f_{\rm RK}(x) dx =
\frac{4\pi A(\kappa,t) S(\kappa,t)}{\Gamma(\kappa+1)} (\kappa t)^{3/2}
 \left( \sum_{i=3}^6 \pi_i(\kappa,t)~
 {\rm B'}\Big(x;\frac{i}{2},\kappa+1-\frac{i}{2},1,\kappa t\Big) \right) R(x)~dx
\label{eq:rkappa_2022}
\end{align}
We also define an auxiliary distribution $F(x)$ without $R(x)$,
\begin{align}
F_{\rm RK}(x) dx =
\frac{4\pi A(\kappa,t) S(\kappa,t)}{\Gamma(\kappa+1)} (\kappa t)^{3/2}
 \left( \sum_{i=3}^6 \pi_i(\kappa,t)~
 {\rm B'}\Big(x;\frac{i}{2},\kappa+1-\frac{i}{2},1,\kappa t\Big) \right) ~dx
.\label{eq:rkappa_envelope}
\end{align}

From Eq.~\eqref{eq:rkappa_envelope}, it is fairly easy to generate the auxiliary distribution $F_{\rm RK}(x)$.
Using a uniform random variate $X_1 \sim U(0,1)$ and the probability $\pi_i(\kappa,t)$,
we choose one of the four beta prime distributions.
Then we obtain $x$ from a random variate according to 
the generalized beta prime distribution, $X_{ {\rm B'}(\cdots) }$.
Using two gamma-distributed random variates and an appropriate scaling (Appendix \ref{sec:beta}),
we obtain
\begin{align}
x
= X_{ {\rm B'}(\frac{i}{2},\kappa+1-i/2,1,\kappa t) }
= \kappa t \times \frac{X_{ {\rm Ga}(i/2,1) }}{X_{ {\rm Ga}(\kappa+1-i/2,1) }}
.
\label{eq:div_rkappa}
\end{align}
Next, we apply the rejection method to $x$. 
Using another uniform random $X_2 \sim U(0,1)$,
if $X_2 < R(x)$, we continue, but 
if not, we repeat the procedure from the beginning.
We similarly use the squeeze technique (Eq.~\eqref{eq:squeeze}),
as discussed in Section \ref{sec:MJ}.
After this, the accepted variates $x$ are distributed by Eq.~\eqref{eq:rkappa_2022}.
We convert $x$ to $p$
\begin{align}
\label{eq:rkappa_x2p}
p = \sqrt{x(x+2)}
\end{align}
and then we further use the spherical scattering (Eq.~\eqref{eq:spherical_scattering}) to obtain the $x$, $y$, and $z$ components of $\vec{p}$.
This $\vec{p}$ follows the relativistic kappa distribution in three-dimensional momentum space (Eq.~\eqref{eq:rkappa}).
The entire procedure is summarized in 
Algorithm 3 in Table~\ref{table:rkappa} in the pseudo code.
Very fortunately, the normalization constant in Eq.~\eqref{eq:rkappa_constant} does not appear in this procedure.
It is interesting to see that
the gamma distributions $Ga(x; i/2,1)$ are divided by different gamma distributions $Ga(x; \kappa+1-i,1)$.

\begin{figure}[htbp]
\centering
\includegraphics[width={\columnwidth}]{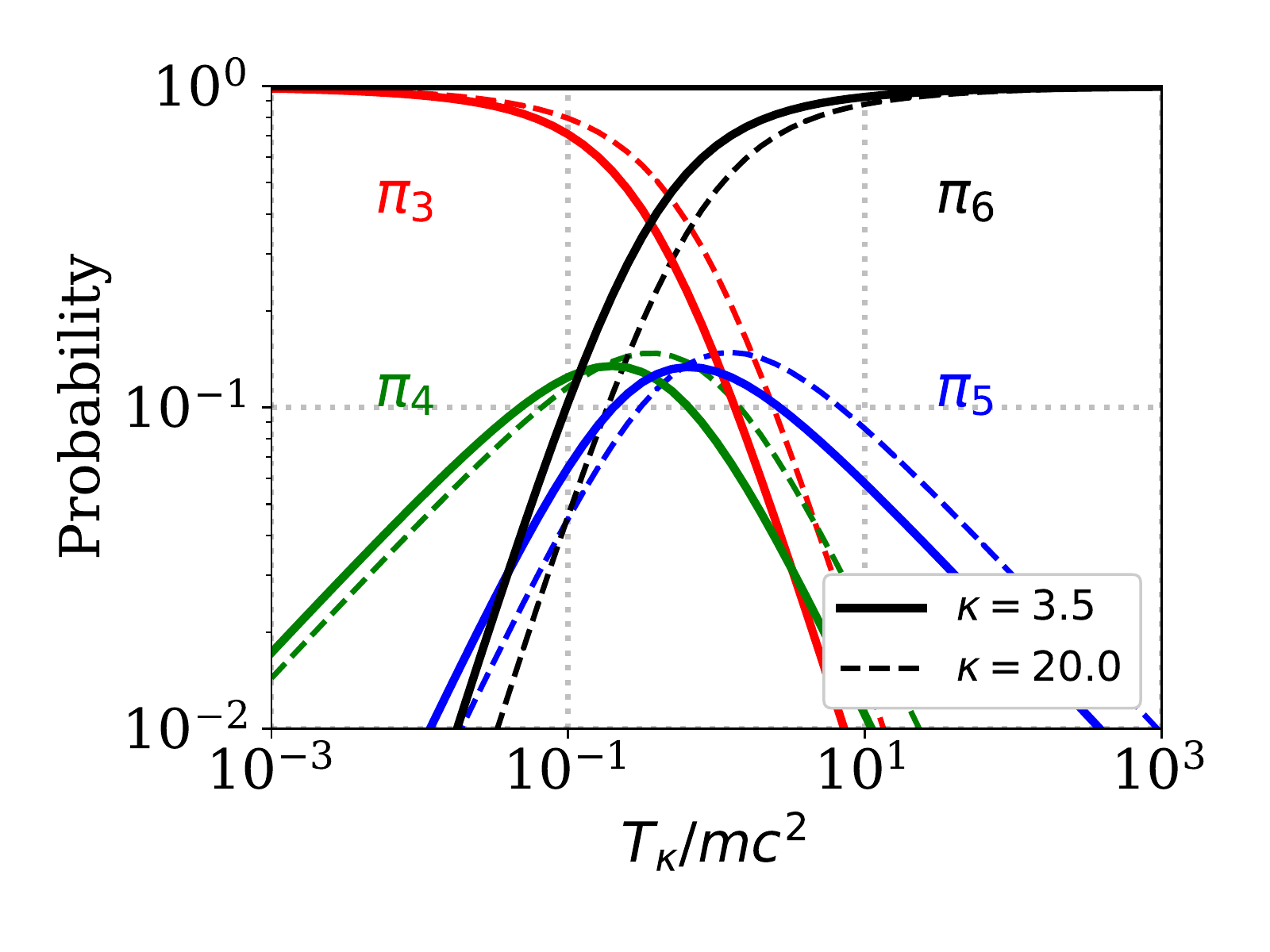}
\caption{
Probability densities $\pi_3, \pi_4, \pi_5, \pi_6$ (Eq.~\eqref{eq:rkappa_prob}) as a function of the characteristic temperature $t=T_{\kappa}/mc^2$. The thick solid lines and the dashed lines indicate the probabilities for $\kappa=3.5$ and for $\kappa=20.0$, respectively.
\label{fig:rkappa_prob}}
\end{figure}

The probability density $\pi_i(\kappa,t)$ is a function of $\kappa$ and $t=T_{\kappa}/mc^2$. Fig.~\ref{fig:rkappa_prob} displays the $t$-dependence of the probability densities for $\kappa=3.5$ and $\kappa=20$,
in the same format as in Fig.~\ref{fig:MJ_prob}.
The $\kappa=20$ case is almost the same as
the \MJ case in Fig.~\ref{fig:MJ_prob}.
This is reasonable, because the $\kappa\rightarrow\infty$ limit corresponds to the \MJ distribution. 
As $\kappa$ decreases, these curves slightly move to the left, but we always see similar trends: the $\pi_3$ component is dominant for $t \ll 1$, the $\pi_4$ and $\pi_5$ components appear in the medium temperature range, and the $\pi_6$ component dominates for $t \gg 1$.
In the nonrelativistic limit of $t \ll 1$,
Eq.~\eqref{eq:rakappa_beta3} is a major component.
The distribution (Eq.~\eqref{eq:rkappa_2022}) approaches $B'(x;3/2;\kappa-1/2,1,\kappa t)$.
Using $\nu=2\kappa-1$, $(1/2)m\theta^2 \simeq t$, and Eq.~\eqref{eq:rkappa_x2p},
we recover $v \simeq (2x)^{1/2} \approx \sqrt{ 2\kappa t X_{{\rm B'}(3/2;\nu/2)} } \simeq \sqrt{ \kappa \theta^2 X_{{\rm B'}(3/2;\nu/2)} }$ in agreement with the nonrelativistic result (Eq.~\eqref{eq:kappa_betap}).
In the ultrarelativistic limit of $t \gg 1$, Eq.~\eqref{eq:rakappa_beta6} dominates.
In this limit, since Eq.~\eqref{eq:rkappa_u2} is approximated by
\begin{align}
f_{\rm RK}(p)dp \propto \Big( 1 + \frac{ p }{\kappa t} \Big)^{-(\kappa+1)} p^2 dp
\label{eq:rela_limit}
\end{align}
and since $p \approx x$,
the distribution approaches ${\rm B'}(3,\kappa-2,1,\kappa t)$ (Eq.~\eqref{eq:rakappa_beta6}).
From Eq.~\eqref{eq:rela_limit}, it is clear that
the energy density $\int_0^{\infty} \gamma f_{\rm RK}(p)dp \propto \int^{\infty}_0 p^{3-\kappa} dp$ diverges to infinity at $\kappa=3$.

\begin{table}
\begin{center}
\caption{An algorithm to generate a relativistic kappa distribution.
$\kappa > 3$ is required.
See Appendix \ref{sec:gamma} for gamma generators.
\label{table:rkappa}}
\begin{tabular}{l}
\\
\hline
{\bf Algorithm 3: New method}\\
\hline
$a \leftarrow 0.56,~~b \leftarrow 0.35,~~R_0 \leftarrow 0.95$\\
compute $\pi_{3},\pi_{4},\pi_{5}$ for given $\kappa,t$ using Eqs. \eqref{eq:rkappa_weight}--\eqref{eq:rkappa_prob}\\
{\bf repeat}\\
~~~~generate $X_1, X_2 \sim U(0, 1)$ \\
~~~~{\bf if}~~~~~~$X_1 < \pi_{3}$ ~~~~~~~~~~~~~~~{\bf then} $i \leftarrow 3$\\
~~~~{\bf elseif} $X_1 < \pi_{3}+\pi_{4}$ ~~~~~~~~{\bf then} $i \leftarrow 4$\\
~~~~{\bf elseif} $X_1 < \pi_{3}+\pi_{4}+\pi_{5}$ ~{\bf then} $i \leftarrow 5$\\
~~~~{\bf else} $i \leftarrow 6$\\
~~~~{\bf endif}\\
~~~~generate $X_3 \sim {\rm Ga}( i/2,1 ), X_4 \sim {\rm Ga}( \kappa + 1 - i/2, 1 )$\\
~~~~$x \leftarrow \kappa t \times \dfrac{ X_3 }{ X_4 }$\\
{\bf until} $X_2 < R_0$ {\bf or} $X_2 < R(x;a,b)$
\\
generate $X_5, X_6 \sim U(0, 1)$ \\
$p \leftarrow \sqrt{ x \left( x+2 \right) }$ \\
$p_x \leftarrow p ~ ( 2 X_5 - 1 )$ \\
$p_y \leftarrow 2 p \sqrt{ X_5 (1-X_5) } \cos(2\pi X_6)$ \\
$p_z \leftarrow 2 p \sqrt{ X_5 (1-X_5) } \sin(2\pi X_6)$ \\
\hline
\end{tabular}
\end{center}
\end{table}

\section{Numerical tests}
\label{sec:tests}

\begin{figure}[htbp]
\centering
\includegraphics[width={\columnwidth}]{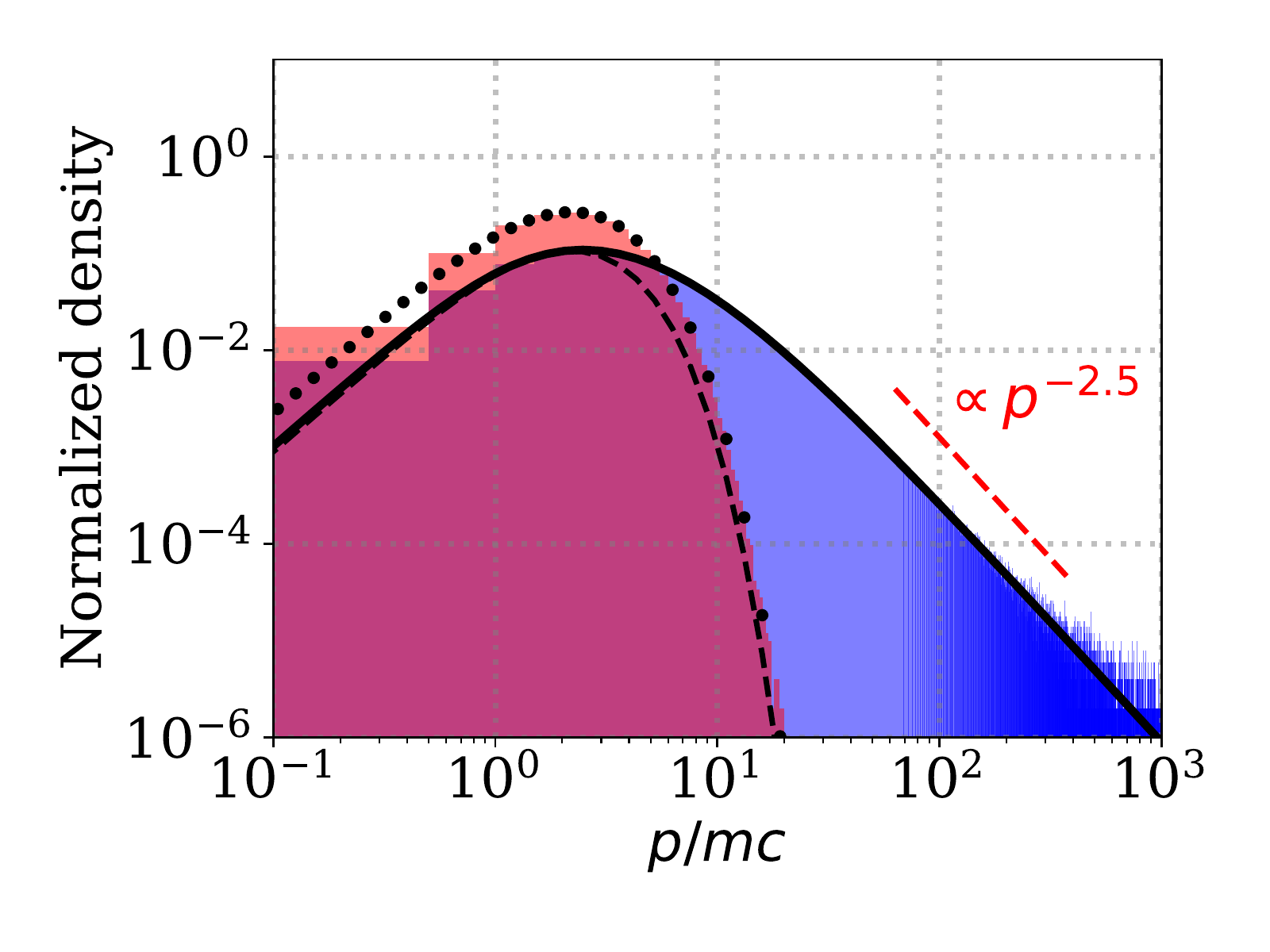}
\caption{
Momentum ($p=m\gamma v$) distribution of the relativistic distributions.
The red histogram and the black dotted line indicate
the distributed particles and the analytic solution
for the \MJ distribution.
The blue histogram and the black solid line indicate
those for the relativistic kappa distribution with $\kappa=3.5$.
The black dashed line indicates the adjusted Maxwellian.
\label{fig:rkappa_u}}
\end{figure}

\begin{figure}[htbp]
\centering
\includegraphics[width={\columnwidth}]{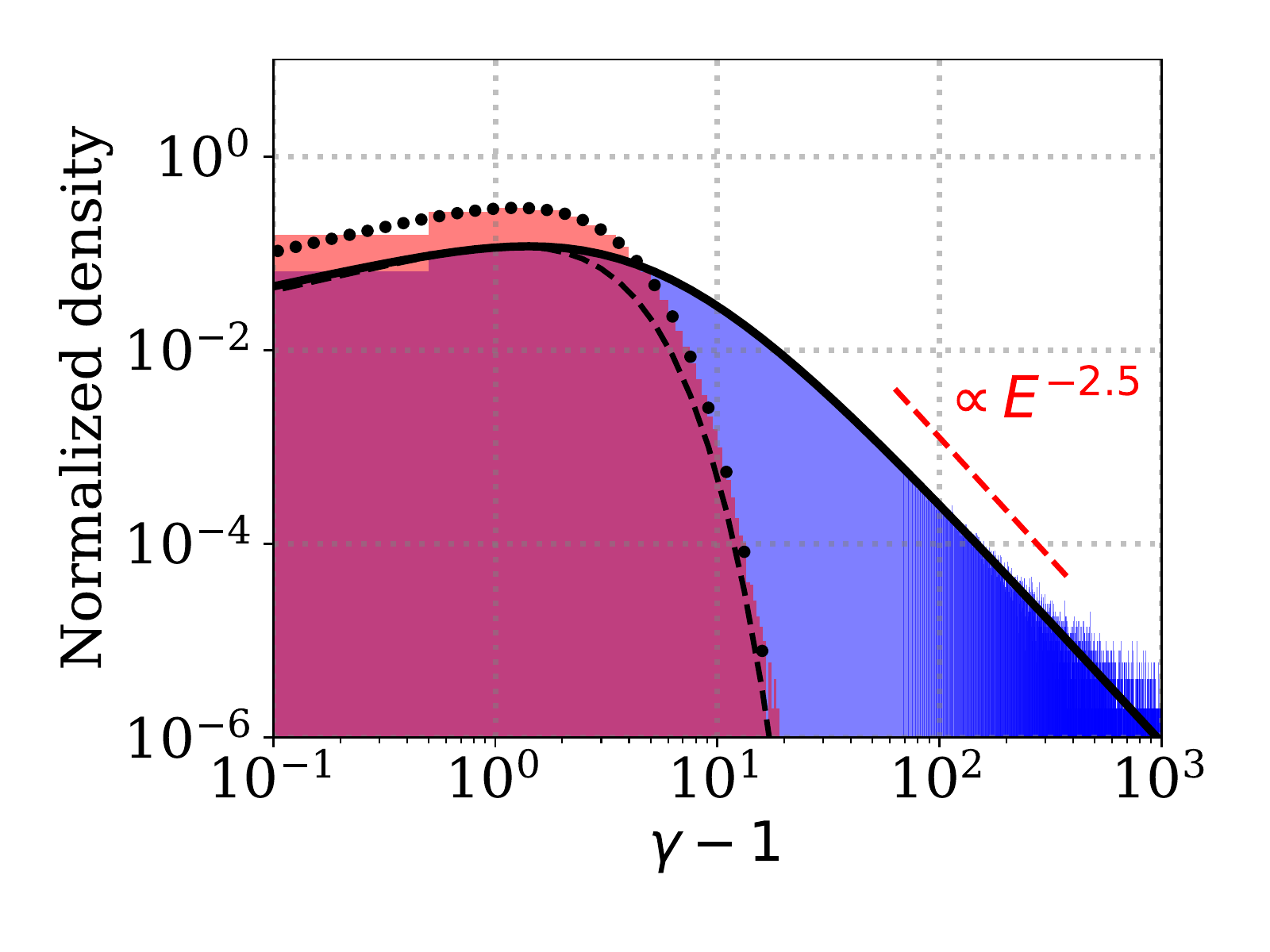}
\caption{
Energy ($\gamma-1$) distribution of the \MJ (the red histogram),
the relativistic kappa distribution with $\kappa=3.5$ (the blue histogram),
and the inscribed Maxwellian (the black dashed line),
in a similar format to Fig.~\ref{fig:rkappa_u}.
\label{fig:rkappa_E}}
\end{figure}

To check our procedures,
we have generated a \MJ distribution with $T_M=1$ and a relativistic kappa distribution with $T_{\kappa}=1$ and $\kappa=3.5$, by using $10^6$ particles.
The modified Canfield method (algorithm 2; Table~\ref{table:MJ}) for the \MJ distribution and
our new method (algorithm 3; Table~\ref{table:rkappa}) for the kappa distribution are used.
In Figures \ref{fig:rkappa_u} and \ref{fig:rkappa_E},
the red histograms show
particles in the \MJ distribution, while
the blue histograms show
particles in the relativistic kappa distribution.
For comparison, theoretical values are indicated by
the dotted and solid lines, respectively.
The numerical results and theoretical curves are in excellent agreement.
In addition, the dashed lines indicate the adjusted \MJ distribution, which is inscribed to the relativistic kappa distribution at the energy of $\gamma-1=T_{\kappa}/mc^2$ or at $p=\sqrt{3}$.
One can clearly separate
the thermal core and the nonthermal tail
by the dashed lines.
Both in Figures \ref{fig:rkappa_u} and \ref{fig:rkappa_E},
the high-energy tail exhibits a power-law distribution. 
As evident in Eq.~\eqref{eq:rela_limit},
the tail scales like $f(p) dp \propto p^{-(\kappa-1)} dp$.
Since $\gamma \approx p$, this also tells
$f(E) dE \propto E^{-(\kappa-1)} dE$.
In this case, the high-energy tail carries $96\%$ of the kinetic energy.
This is substantially higher than in the nonrelativistic distribution with the same kappa index.
This further supports the importance of the numerical algorithm that precisely reproduces the high-energy tail.

\begin{figure}[htbp]
\centering
\includegraphics[width={\columnwidth}]{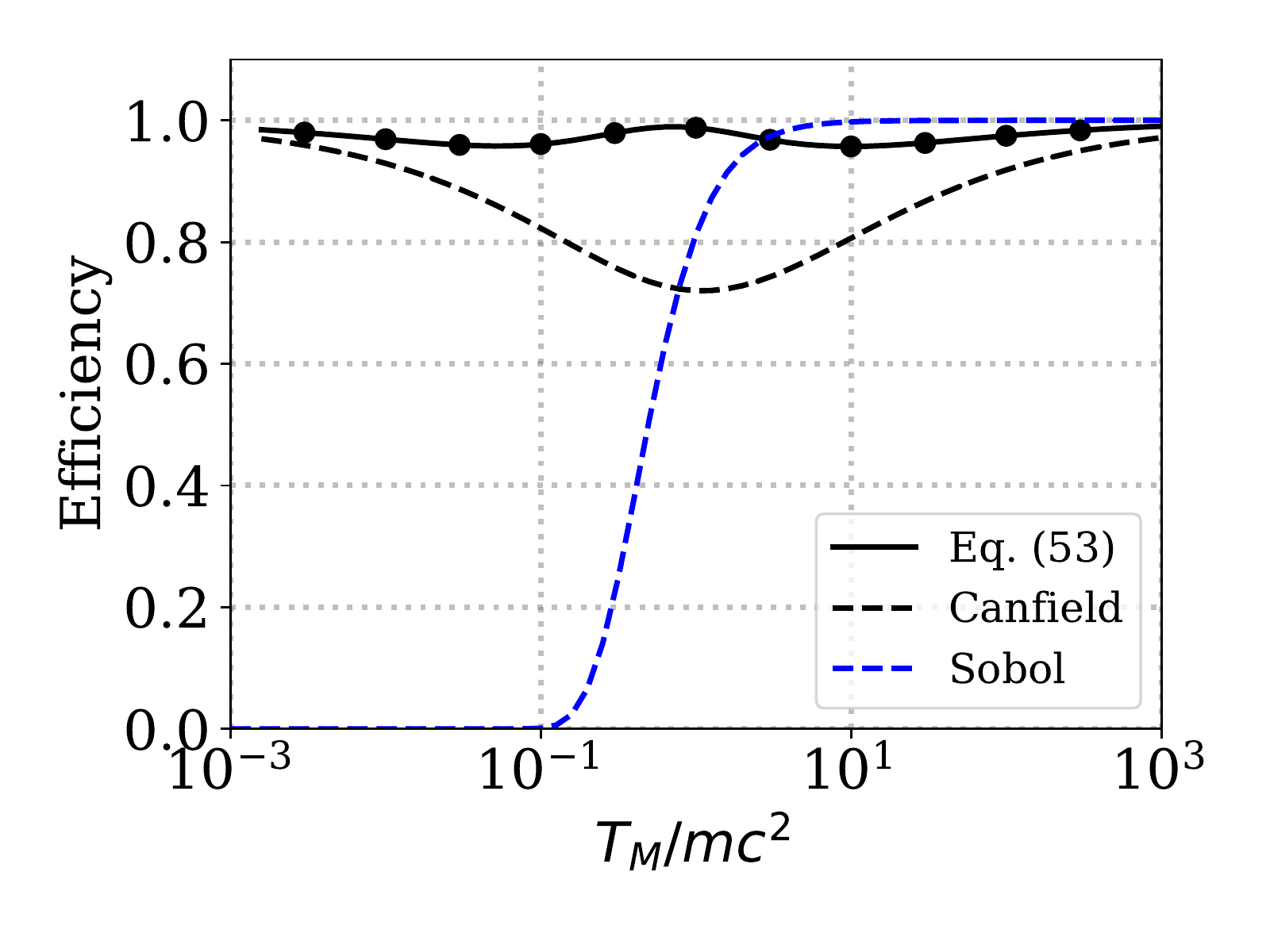}
\caption{
Efficiency of the rejection methods for the \MJ distribution.
The black circles indicate our numerical results,
the black line shows the efficiency of our method (Eq.~\eqref{eq:MJ_eff}),
the black dashed line shows the efficiency of the Canfield method (Eq.~\eqref{eq:MJ_eff} with $a=b=1$),  and
the blue line shows that of
the Sobol method (Eq.~\eqref{eq:sobol_eff}).
\label{fig:MJ_eff}}
\end{figure}

Since the relativistic algorithms rely on the rejection method, we have evaluated their acceptance efficiency,
by using $10^6$ particles.
We first evaluate the modified Canfield method.
Fig.~\ref{fig:MJ_eff} show
the efficiency as a function of $t=T_M/mc^2$.
The closed circles indicate our numerical results.
We find very good efficiency of $\gtrsim 95 \%$ regardless of $t$.
To obtain a distribution of Eq.~\eqref{eq:canfield87},
we first generate the auxiliary distribution with Eq.~\eqref{eq:envelope} and then apply the rejection.
Therefore the acceptance rate can be estimated by
\begin{align}
{\rm eff}_{\rm MJ}(t; a,b) &= 
\frac{
\int_0^{\infty} f_{\rm MJ}(x) dx
}{
\int_0^{\infty} F_{\rm MJ}(x) dx 
}
=
\dfrac{
1
}{
\dfrac{\sqrt{t}e^{-1/t} S(t) }{\sqrt{2} K_2 (1/t)}
\sum_{i=3}^6 \left\{ \pi_i(t) \int_0^{\infty} Ga(x) dx \right\}
}
\nonumber \\
&=
\label{eq:MJ_eff}
\frac{\sqrt{2} e^{1/t} K_2 (1/t)}
{\sqrt{t}
\left( \sqrt{\pi} + a\cdot\sqrt{2t} + b\cdot\frac{3\sqrt{\pi}t}{2} + {(2t)^{3/2}}
\right)}
.
\end{align}
This is indicated by the black curve in Fig.~\ref{fig:MJ_eff},
in excellent agreement with the numerical results.

For comparison, we show
the efficiency of the original Canfield method (the black dashed line) and the Sobol method\citep{sobol76} (the blue dashed line).
We estimate the efficiency of the Canfield method
by substituting $(a,b)=(1,1)$ in Eq.~\eqref{eq:MJ_eff}.
It has the minimum is $71.99 \%$ at $t \approx 1.1$ and
it is asymptotic to 1 in the $t \ll 1$ and $t \gg 1$ limits.
The curves of the original and modified Canfield methods are similar to,
but not identical to, the rejection functions in Fig.~\ref{fig:rej}.
This is reasonable, because
the rejection function near the most probable speed 
highly controls the overall efficiency.
In addition, the Sobol method is a popular rejection-based algorithm,
which is based on the third-order gamma distribution.
Its efficiency is given by \citep{poz83}
\begin{equation}
\label{eq:sobol_eff}
{\rm eff}_{\rm Sobol}(t)
=
\frac{1}{2 t^2} K_2(1/t)
.
\end{equation}
As can be seen in Fig.~\ref{fig:MJ_eff},
the Sobol method is superior for $t\gtrsim 3$, but
the modified Canfield method also gives good results.
In contrast, although the Sobol method becomes notoriously inefficient for $t \lesssim 1$,
the other methods remain efficient.
In short, the modified Canfield method outperforms the other two for $t \lesssim 1$ and it is as competitive as the Sobol method for $t \gtrsim 3$.

\begin{figure}[htbp]
\centering
\includegraphics[width={\columnwidth}]{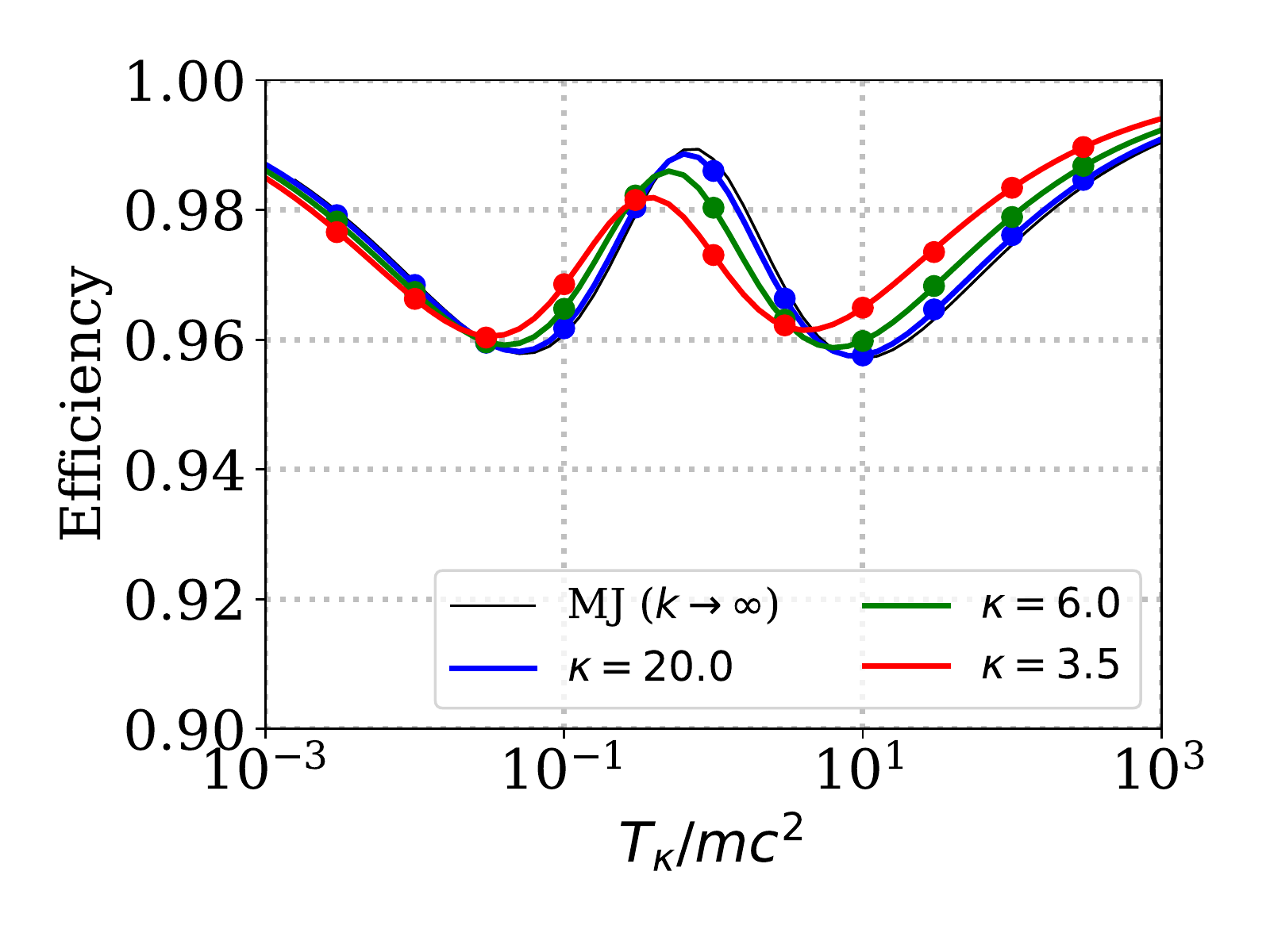}
\caption{
Efficiency of the rejection method for the relativistic kappa distribution.
The closed circles indicate our numerical results.
The color solid lines show theoretical values by Eq.~\eqref{eq:rkappa_eff}.
The black line indicates a theoretical curve for the \MJ distribution (Eq.~\eqref{eq:MJ_eff}).
\label{fig:rkappa_eff}}
\end{figure}

Next, we discuss the acceptance efficiency of our new method for the relativistic kappa distribution. 
Fig.~\ref{fig:rkappa_eff} presents the efficiency
as a function of $t=T_{\kappa}/mc^2$.
The closed circles indicate our numerical results with $10^6$ particles, and the solid lines are the theoretical curves, for $\kappa=3.5$, $6.0$, and $20.0$.
We similarly compare the auxiliary function (Eq.~\eqref{eq:rkappa_envelope}) and the distribution (Eq.~\eqref{eq:rkappa_2022}) 
to estimate the theoretical values,
\begin{align}
{\rm eff}_{\rm RK}(\kappa,t)
&=
\frac{
\int_0^{\infty} f_{\rm RK}(x) dx
}{
\int_0^{\infty} F_{\rm RK}(x) dx 
}
=
\dfrac{1}
{
\dfrac{4\pi A(\kappa,t) S(\kappa,t)
(\kappa t)^{3/2}}
{\Gamma(\kappa+1)}
\sum_{i=3}^6 \left\{ \pi_i(t) \int_0^{\infty} {\rm B'}(x) dx \right\}
}
\nonumber \\
&=
\frac{
{(\pi)^{1/2}}
~\Gamma\left( \kappa + 2 \right)
\Gamma\left( \kappa - 2 \right)
~{}_2F_1\left( -\frac{3}{2}, \frac{5}{2}; \kappa+\frac{1}{2}; 1-\frac{\kappa t}{2}\right)
}
{
\Gamma\left( \kappa +  \frac{1}{2} \right) \Big(
\sqrt{\pi}
{\Gamma(\kappa-\frac{1}{2})}
+
a\cdot\sqrt{2\kappa t}~
\Gamma(\kappa-1)
+
b\cdot\frac{3\sqrt{\pi}}{2}
{(\kappa t)~
\Gamma(\kappa-\frac{3}{2})}
+
{(2\kappa t)^{3/2}~
\Gamma(\kappa-2)}
\Big)
}
. \label{eq:rkappa_eff}
\end{align}
For reference, the theoretical curve for the \MJ distribution (Eq.~\eqref{eq:MJ_eff}) is also plotted by the black line.
We find excellent efficiencies of $\gtrsim 95\%$ in all the cases. In particular, it always exceeds $96\%$ for $\kappa=3.5$.
As $\kappa$ increases, the curve gradually approaches that of the \MJ case, because the kappa distribution approaches the \MJ distribution in the $\kappa\rightarrow \infty$ limit.
As $\kappa$ decreases,
locations of the local peak and the local minima
move to the left direction.
This can be interpreted as follows.
As evident in Figs.~\ref{fig:rkappa_u} and \ref{fig:rkappa_E},
as $\kappa$ decreases,
the relativistic kappa distribution contains more particles in the high-energy tail and less particles in the low-energy part than the \MJ distribution with the same temperature $t$.
Therefore the distribution interacts with the rejection function similarly to the Maxwell distributions with higher-temperature. 
This is consistent with the fact that
the probabilities are shifted to the left in Fig.~\ref{fig:rkappa_prob}.
Aside from this difference,
the overall efficiency exceeds $\gtrsim 95\%$ in all cases.
This is an outcome of our new rejection function (Eq.~\eqref{eq:rkappa_rej}).

Technically, it is possible to generate the relativistic kappa distribution in a similar way to the original Canfield method ($a=b=1$).
As far as we have tested,
the efficiency drops to $72$--$74 \%$ around $t=T_{\kappa}/mc^2 \sim 1$.
This is also expected from the comparison for the \MJ distribution in Fig.~\ref{fig:MJ_eff}.
Since the modified Canfield method is always more efficient than the original Canfield method, we find no reason to employ $a=b=1$.
We recommend the readers to use our optimum parameter $(a,b)=(0.56, 0.35)$ for kappa distributions. 

\section{Discussion and summary}
\label{sec:discussion}

The proposed methods will be used at startup in PIC or Monte Carlo simulations. 
The algorithms do not need to be the fastest,
but it is important to know their computational cost.
As an indicator of the cost,
we estimate the number of random variates per particle in the procedures.
To generate a nonrelativistic Maxwellian,
we need 3 normal random variates per particle.
For the \MJ distribution,
according to Tables~\ref{table:MJ} and \ref{table:gamma},
the modified Canfield method uses 3--5 uniform or normal variates in the rejection loop and another 2 uniform variates outside the loop.
Considering ${\rm eff}_{\rm MJ} \approx 1$,
it typically uses $\lesssim 7$ variates per particle.
For the relativistic kappa distribution, the number depends on $\kappa$.
Judging from Tables~\ref{table:rkappa} and \ref{table:gamma},
our method uses $(\kappa+4)$ random variates in the loop in the worst case,
and another 2 outside the loop.
Considering ${\rm eff}_{\rm RK} \approx 1$,
the number is $\lesssim (\kappa + 6)$ per particle.
Though the numerical cost will increase for higher $\kappa$,
it is still predictable.
For another examples,
if we employ another gamma generator in Table~\ref{table:gamma3},
and if we assume its acceptance efficiency to be $\approx 1$,
the modified Canfield method uses $\approx 6$ uniform or normal variates per particle
for the \MJ distribution.
Similarly, the new kappa method uses $\approx 8$ variates,
regardless of $\kappa$. 
The numerical cost can be smaller. 
Considering these issues,
the numerical cost is unlikely to be a big problem.

We remark that
the $\kappa$ parameter should satisfy $\kappa > 3$ in our method.
Even though the nonrelativistic kappa distribution is defined for
$\kappa > 3/2$,
our method does not permit $3/2 < \kappa \le 3$
in the nonrelativistic limit. 
This is because our method uses
the ${\rm B'}(3,\kappa-2,1,\kappa t)$ component (Eq.~\eqref{eq:rakappa_beta6}),
whose energy increases to infinity at $\kappa=3$.
To relax this constraint,
one option is to drop the ${\rm B'}(x; 3,\kappa-2,1,\kappa t)$ component in the nonrelativistic regime.
In such a case, the next ${\rm B'}(x; 5/2,\kappa-3/2,1,\kappa t)$ component
requires $\kappa > 5/2$.
To justify this, the probability $\pi_6$ (Eq.~\eqref{eq:rkappa_prob}) needs to be small enough, however, it is non-negligible (Fig.~\ref{fig:rkappa_prob}). 
For example, $\pi_6(3.5,10^{-2})\approx 4\times 10^{-3}$ and
$\pi_6(3.5,10^{-3})\approx 10^{-4}$.
Therefore, this option will be useful only in a limited case of $t \ll 1$.

Although it is unlikely,
when we use a very large number of particles,
the denominators in Eqs.~\eqref{eq:div_rkappa} or Eq.~\eqref{eq:v_recovery} may be zero.
To avoid the division by zero, we have three options.
First, we can simply check the denominator, and then
regenerate the variate if necessary.
Second, we may employ a gamma generator that never return zero.
For example, if uniform variates are defined in the interval $(0,1)$, then the gamma generator in Table \ref{table:gamma} does not return zero. 
For another example, at least theoretically,
the gamma generator in Table \ref{table:gamma3} does not return zero.
Third, we may add a very small number $\epsilon$
to the denominators in Eqs.~\eqref{eq:div_rkappa} and \eqref{eq:v_recovery}.
The gamma distribution ${\rm Ga}(x;k,\lambda)$ starts at the origin when $k>1$, and the mode (the most frequent value) of the distribution $x_m \equiv \lambda(k-1)$ can be regarded as a typical value of $X_{\rm Ga (k,\lambda)}$.
In such a case, it would be plausible to set
$\epsilon$ to an order of a machine epsilon of $x_m$, i.e.,
$10^{-12} \times \mathcal{O}(x_m)$ in double precision and
$10^{-7} \times \mathcal{O}(x_m)$ in single precision.
When $k\rightarrow 1$ or $k = 1$,
population near $x=0$ is non-negligible.
This occurs when $\kappa \rightarrow 3$ in the relativistic kappa distribution and
$\kappa \rightarrow 3/2$ in the nonrelativistic kappa distribution.
Simulation settings need to be reconsidered,
because the kappa distributions cannot be defined in these limits.

When plasmas are drifting, we need to consider the Lorentz transformation. 
In order to transform a velocity distribution function,
it is necessary to transform the volume element,
as well as the particle energy and momentum.
This often requires another rejection process.
In such a case, we recommend the readers to combine
the procedures in this paper and
the volume transform methods in Section III in Ref.~\onlinecite{zeni15b}. 
Finally, we remark that
our methods essentially take advantage of
the following transform of an isotropic distribution,
\begin{align}
p^2dp \rightarrow
\Big( \sqrt{2}{x}^{1/2} + a{x} + b\sqrt{2}{x}^{3/2}+{x}^2 \Big)
R(x;a,b)
~ dx,
\nonumber
\end{align}
and we show $R(x;a,b)\gtrsim 95\%$ for $(a,b)=(0.56, 0.35)$ in Fig.~\ref{fig:rej}.
Thus, as long as isotropic,
it might be possible to apply our efficient rejection function ($\gtrsim 95\%$)
to other relativistic velocity distributions.

In summary, we have presented a rejection-based algorithm
for generating the relativistic kappa distributions.
As preparation, we have briefly reviewed
algorithms for the nonrelativistic distributions.
Then we have presented
the modified Canfield method for the \MJ distribution.
We have improved its rejection function,
by introducing the two hyperparameters.
Using this and generalized beta prime distributions,
we have constructed a rejection-based algorithm,
presented in Table~\ref{table:rkappa}.
These procedures are numerically tested.
The new method excellently generates the distribution,
including its power-law tail.
For \MJ distribution, we have improved the acceptance rate
from $\gtrsim 72\%$ to $\gtrsim 95\%$.
The new method for the relativistic kappa distribution
takes advantage of the efficient rejection function.
Its efficiency is $\gtrsim 95\%$
from the nonrelativistic to ultrarelativistic temperature regimes.
The algorithms use a predictable number of random variates.
The entire procedure is simple enough to avoid bugs.
Based on these facts, our algorithm is practically useful.
We hope that these algorithms are useful in PIC and Monte Carlo simulations in relativistic plasma physics.

\begin{acknowledgements}
One of the authors (SZ) acknowledges G. R. Werner for stimulating discussion. This work was supported by Grant-in-Aid for Scientific Research (C) 21K03627 and (S) 17H06140 from the Japan Society for the Promotion of Science (JSPS).
\end{acknowledgements}

{\bf Data Availability}

The data supporting the findings of this study are available from the corresponding author upon reasonable request.

\appendix

\section{Generating gamma distributions}
\label{sec:gamma}

There are several algorithms to generate
a random variate according to the gamma distribution $Ga(x; k, \lambda)$.\citep{devroye86,mt00,kroese11,yotsuji10}
One can use gamma generators provided by software packages, or
one can write one's own generators to make the code portable.
For the latter purpose, we outline algorithms to load the gamma distributions.

The gamma distribution with an integer parameter $k$,
also known as Erlang distribution,
can be generated by using uniform random variates $U_{i} \sim U(0,1)$
\begin{align}
\label{eq:erlang}
X_{{\rm Ga}(k,\lambda)} &= -\lambda \ln \prod_{i=1}^k U_{i}
\end{align}
For a half integer $k$, one can generate the distribution by using $[k]$ independent uniform random variates and one normal random variate $n_1 \sim \mathcal{N}(0,1)$,\citep{devroye86}
\begin{align}
X_{{\rm Ga}(k,\lambda)}
&=
-\lambda \ln \prod_{i=1}^{[k]}  U_{i} + \left(\frac{\lambda}{2}\right) n_1^2
,
\end{align}
where $[x]$ is the floor function.
The procedure is listed in Table.~\ref{table:gamma}.
Using \BM transform, one can replace the last term $(\lambda/2) n^2 = -\lambda (\ln u_1) \cos^2 2\pi u_2$ by using two additional uniform random variates $u_1$ and $u_2$.
Practically, when the uniform random variate $U_i$ is defined in the interval $[0, 1)$,
it is useful to redefine $U_i = 1-U_i$ to avoid $\log 0$.

For $k>1$, Marsaglia's rejection algorithm is widely used.\citep{mt00,kroese11,yotsuji10}
Here we just display the procedure in Table \ref{table:gamma3}.
For more detail, the readers may wish to consult
the original article\citep{mt00} or recent textbooks on Monte Carlo methods.\citep{yotsuji10}
At the end of each iteration loop, one can see two conditions.
The first condition is optional,
because it is added by the squeeze method.
According to the literature and our tests,
the acceptance efficiency is usually high, $\gtrsim 0.95$.


\begin{table}
\begin{center}
\begin{tabular}{l}
\\
\hline
{\bf Algorithm A}\\
\hline
{\bf function} Gamma-generator($k, \lambda$)\\
generate uniform random $U_1, U_2, \cdots, U_{[k]} \in (0, 1]$ \\
{\bf if} $k$ is an integer {\bf then}\\
~~~~$x \leftarrow - \ln \big( \prod_{i=1}^{k} U_{i} \big)$ \\
{\bf elseif} $k$ is a half integer {\bf then}\\
~~~~generate $n \sim \mathcal{N}(0,1)$ \\
~~~~$x \leftarrow - \ln \big(\prod_{i=1}^{[k]} U_{i} \big) + \dfrac{1}{2} n^2$ \\
{\bf endif}\\
{\bf return} $\lambda x$\\
\hline
\end{tabular}
\caption{Algorithms to generate gamma distributions for an integer or a half-integer parameter $k$. $[x]$ is the floor function.
\label{table:gamma}}
\end{center}
\end{table}

\begin{table}
\begin{center}
\begin{tabular}{l}
\\
\hline
{\bf Algorithm B: Marsaglia's rejection method}\\
\hline
{\bf function} Gamma-generator2($k, \lambda$)\\
$d \leftarrow k - 1/3,~~c \leftarrow 1/\sqrt{9d}$ \\
{\bf repeat}\\
~~~~{\bf repeat}\\
~~~~~~~~generate $x \sim \mathcal{N}(0, 1)$ \\
~~~~~~~~$v \leftarrow 1 + cx$\\
~~~~{\bf until} $v > 0$\\
~~~~generate $u \sim U(0, 1)$ \\
~~~~$v \leftarrow v^3$\\
{\bf until} $1 -0.0331 x^4 > u$ {\bf or} \\
~~~~~~~~~~~~~~~$0.5x^2 + d - dv + d\log v \ge \log u$\\
{\bf return} $\lambda dv$\\
\hline
\end{tabular}
\caption{Marsaglia's algorithm to generate gamma distributions for arbitrary $k>1$.
\label{table:gamma3}}
\end{center}
\end{table}

\section{Generating beta prime distributions}
\label{sec:beta}

We consider two independent random variates $X \sim {\rm Ga}(\alpha,\delta)$ and $Y \sim {\rm Ga}(\beta,\delta)$.
Using them, we consider the following variables,
\begin{align}
U = \frac{X}{Y},~~~V = X + Y
\end{align}
$X$ and $Y$ are expressed as follows
\begin{align}
X = \frac{UV}{1+U}, ~~~Y = \frac{V}{1+U}
\end{align}
The joint probability distribution function of $X, Y$ is
\begin{align}
\label{eq:joint_xy}
f_{X,Y}(x,y) &= 
\frac{1}{\Gamma(\alpha)\delta^{\alpha}}
x^{\alpha-1}
e^{-x/\delta}
\frac{1}{\Gamma(\beta)\delta^{\beta}}
y^{\beta-1}
e^{-y/\delta}
\end{align}
We rewrite this function in terms of $(u,v$)
\begin{align}
f_{U,V}(u,v)
&=
\frac{1}{\Gamma(\alpha)\delta^{\alpha}}
x^{\alpha-1}
e^{-x/\delta}
\frac{1}{\Gamma(\beta)\delta^{\beta}}
y^{\beta-1}
e^{-y/\delta}
\left\| \frac{\partial(x,y)}{\partial(u,v)} \right\|
\nonumber \\
&=
\frac{1}{\Gamma(\alpha)\Gamma(\beta)\delta^{\alpha+\beta}}
\left(\frac{uv}{1+u}\right)^{\alpha-1}
\left(\frac{v}{1+u}\right)^{\beta-1}
e^{-v/\delta}
\frac{v}{(1+u)^2}
\nonumber \\
&=
\left\{
\frac{\Gamma(\alpha+\beta)}{\Gamma(\alpha)\Gamma(\beta)}
u^{\alpha-1}
(1+u)^{-(\alpha+\beta)}
\right\}
\times
\left\{
\frac{v^{(\alpha+\beta)-1} e^{-v/\delta}}{\Gamma(\alpha+\beta)\delta^{\alpha+\beta}}
\right\}
\end{align}
This means that
$u$ and $v$ follow a beta prime distribution $U \sim {\rm B'}(\alpha,\beta)$ and a gamma distribution $V \sim {\rm Ga}(\alpha+\beta,\delta)$, respectively.
This also indicates that $U$ and $V$ are independent.
In other words, the beta prime distribution can be generated from the two gamma distributions.

\section{Inscribed Maxwell distributions for the nonrelativistic kappa distribution}
\label{sec:in_kappa}

We derive a Maxwell distribution which is inscribed to the kappa distribution.
We consider the density ratio of the Maxwellian to the kappa distribution $a(v)$ as a function of $v$. In addition, we consider its two derivatives.
\begin{align}
a(v) &\equiv
\frac{ f_{\rm M}(v)}{ f_{\rm \kappa}(v)}
=
\frac{N_M \Gamma{\left(\kappa - 1/2 \right)} }{N_{\kappa} \Gamma{\left(\kappa + 1 \right)}} 
\left(  \frac{\kappa \theta^2}{v_M^{2}} \right)^{3/2}
\left( 1 + \frac{v^{2}}{\kappa \theta^{2}} \right)^{\kappa + 1}
\exp \left(- \frac{v^{2}}{v_M^{2}} \right)
\label{eq:a0}\\
a'(v) &= 2 a(v) \left\{ v \left[ \frac{ \kappa+1 }{ v^2 + \kappa\theta^{2} } - \frac{1}{v_M^{2}}  \right] \right\}
\label{eq:a1}
\\
a''(v) &= 2 \left\{ \Big( a'(v) v + a(v) \Big) \left[ \frac{ \kappa+1 }{ v^2 + \kappa\theta^{2} } - \frac{1}{v_M^{2}}  \right] -  a(v) \frac{ 2 ( \kappa+1) v^2 }{ (v^2 + \kappa\theta^{2})^2 } \right\}
\end{align}
When the Maxwellian core (Eq.~\eqref{eq:maxwell}) is inscribed to the kappa distribution at $v=v_0$,
the following conditions need to be satisfied:
\begin{align}
\label{eq:kappa_adjust_condition}
a(v_0) &=1, ~~~
a'(v_0) = 0, ~~~
a''(v_0) \le 0,
\end{align}
Since Eq.~\eqref{eq:a1} changes its sign only once,
$v=v_0$ is the only contact point.
The conditions can be simplified to
\begin{align}
v_0 \left[ \frac{ \kappa+1 }{ v_0^2 + \kappa\theta^{2} } - \frac{1}{v_M^{2}}  \right] = 0,
~~
\left( \left[ \frac{ \kappa+1 }{ v_0^2 + \kappa\theta^{2} } - \frac{1}{v_M^{2}}  \right] -  \frac{ 2 ( \kappa+1) v_0^2 }{ (v_0^2 + \kappa\theta^{2})^2 } \right) 
\le 0,
\end{align}
and then we obtain the following solution $v_M$
\begin{align}
\label{eq:vm}
v_M^2 =
\frac{1}{ \kappa+1 } \left( v_0^2 + \kappa\theta^{2} \right)
.
\end{align}

There can be several choices for $v_0$.
If we employ $v_0=0$, then Maxwellian parameters are given by
\begin{align}
v_M^2 = \frac{\kappa}{\kappa+1}\theta^{2},~~~
N_M = \left( \frac{1}{\kappa+1}\right)^{3/2} \frac{\Gamma(\kappa+1)}{\Gamma(\kappa-1/2)} N_{\kappa}.
\end{align}
This is identical to the asymptotic Maxwellian core in the $v \rightarrow 0$ limit
~[for example, Eq.~(12) in \citet{livadiotis09}].
If we employ the most probable speed $v_0=\theta$,
then we obtain \citet{oka13}'s adjusted Maxwellian
\begin{align}
v_M = \theta,~~~
N_M = e \frac{\Gamma(\kappa+1)}{\Gamma(\kappa-1/2)} \kappa^{-\frac{3}{2}} \bigg( 1+\frac{1}{\kappa} \bigg)^{-(\kappa+1)} N_{\kappa},
\end{align}
where $e$ is Euler's number.
This adjusted Maxwellian is useful to evaluate
the thermal and nonthermal parts of the kappa distribution.\citep{oka13}

\section{Inscribed \MJ distributions for the relativistic kappa distribution}
\label{sec:in_rkappa}

We derive a \MJ distribution which is inscribed to the relativistic kappa distribution.
We consider the ratio function $a(E)$, as a function of the particle energy $E$,
\begin{align}
a({E})
&\equiv
\dfrac{f_{\rm MJ}(E)}{f_{\rm RK}(E)}
=
\dfrac{N_M e^{-1/T_M} }{4\pi A(\kappa,T_{\kappa}) T_M K_2\Big(\frac{1}{T_M}\Big) } \Big( 1 + \frac{ E }{\kappa T_{\kappa}} \Big)^{\kappa+1} 
\exp\left(-\frac{E}{T_M}\right)
\label{eq:rel_a0}\\
a'({E})
&=
a(E)
\left[
\frac{\kappa+1}{\kappa T_{\kappa}+E}
-
\frac{1}{T_M}
\right]
\label{eq:rel_a1}\\
a''(E)
&=
a(E)
\left\{
\left[
\frac{\kappa+1}{\kappa T_{\kappa}+E}
-
\frac{1}{T_M}
\right]^2
-
\frac{\kappa+1}{(\kappa T_{\kappa}+E)^2}
\right\}
\end{align}
When the \MJ distribution is inscribed to the relativistic kappa distribution
at the energy of $E_0$, the following conditions need to be satisfied,
\begin{align}
a(E_0) = 1,~~~a'(E_0) = 0,~~~a''(E_0) < 0.
\end{align}
Since Eq.~\eqref{eq:rel_a1} changes its sign only once,
$E=E_0$ is the unique contact point.
These conditions lead to the following solution,
\begin{align}
T_M &= \frac{\kappa T_{\kappa} + E_0}{\kappa+1}.
\end{align}

We have several choices for $E_0$.
For $E_0=0$, we obtain the following solutions.
Note that $A(\kappa,t)$ contains $N_{\kappa}$.
\begin{align}
T_M &= \frac{\kappa}{\kappa+1} T_{\kappa},~~~
N_M = 4\pi A(\kappa,t) T_M K_2 \Big(\frac{1}{T_M} \Big)  e^{1/T_M}
\end{align}
For $E_0=T_{\kappa}$, we obtain a relativistic extension of \citet{oka13}'s adjusted Maxwellian.
\begin{align}
T_M &= T_{\kappa},~~~
N_M = {4\pi e} A(\kappa,t) T_{\kappa} K_2 \Big(\frac{1}{T_{\kappa}} \Big) e^{1/T_{\kappa}} \Big( 1 + \frac{ 1 }{\kappa} \Big)^{-(\kappa+1)}
\end{align}

\section{Finding optimum hyperparameters}
\label{sec:grid}

\begin{figure*}[htbp]
\centering
\includegraphics[width={0.49\textwidth}]{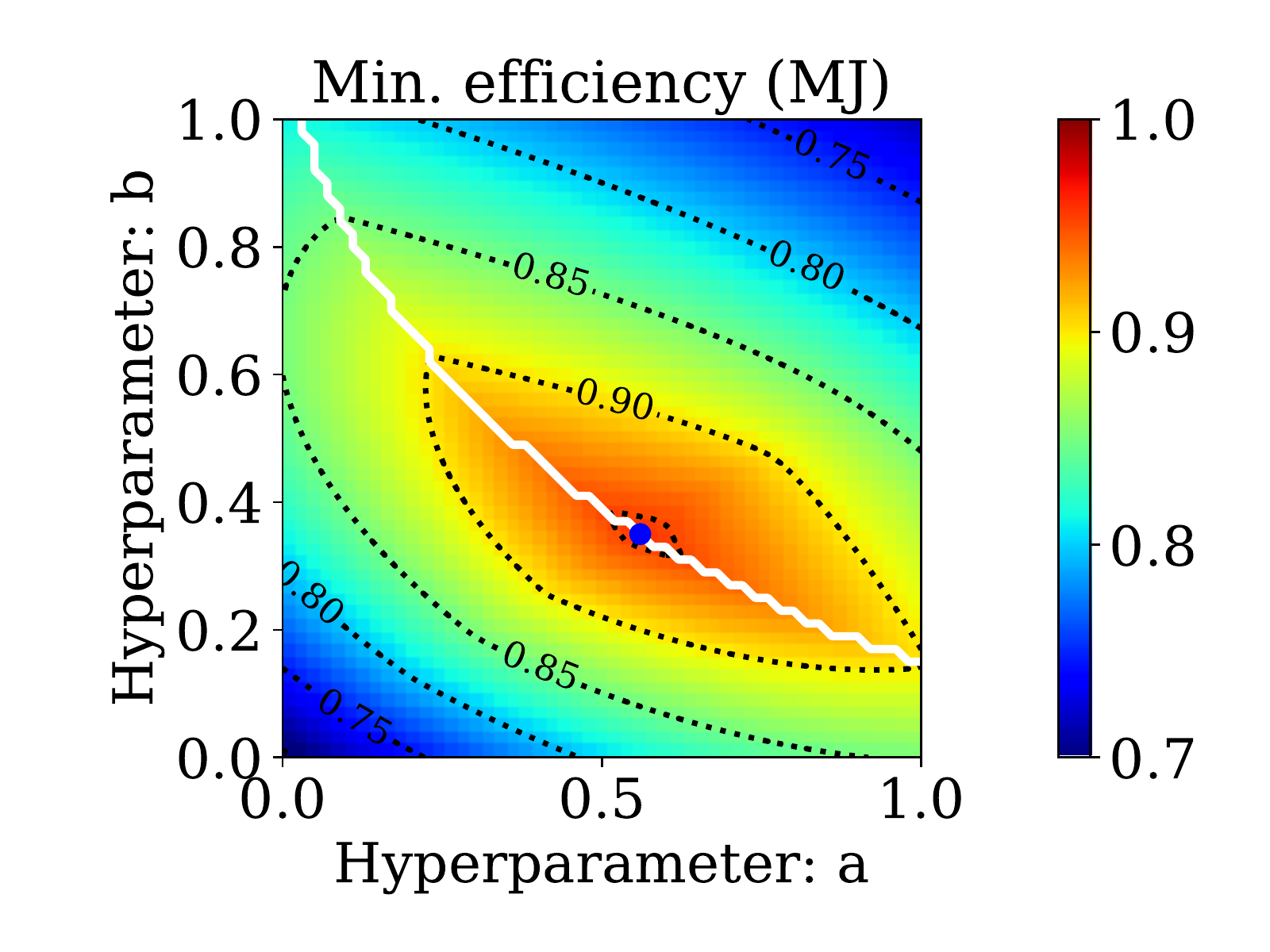}
\includegraphics[width={0.49\textwidth}]{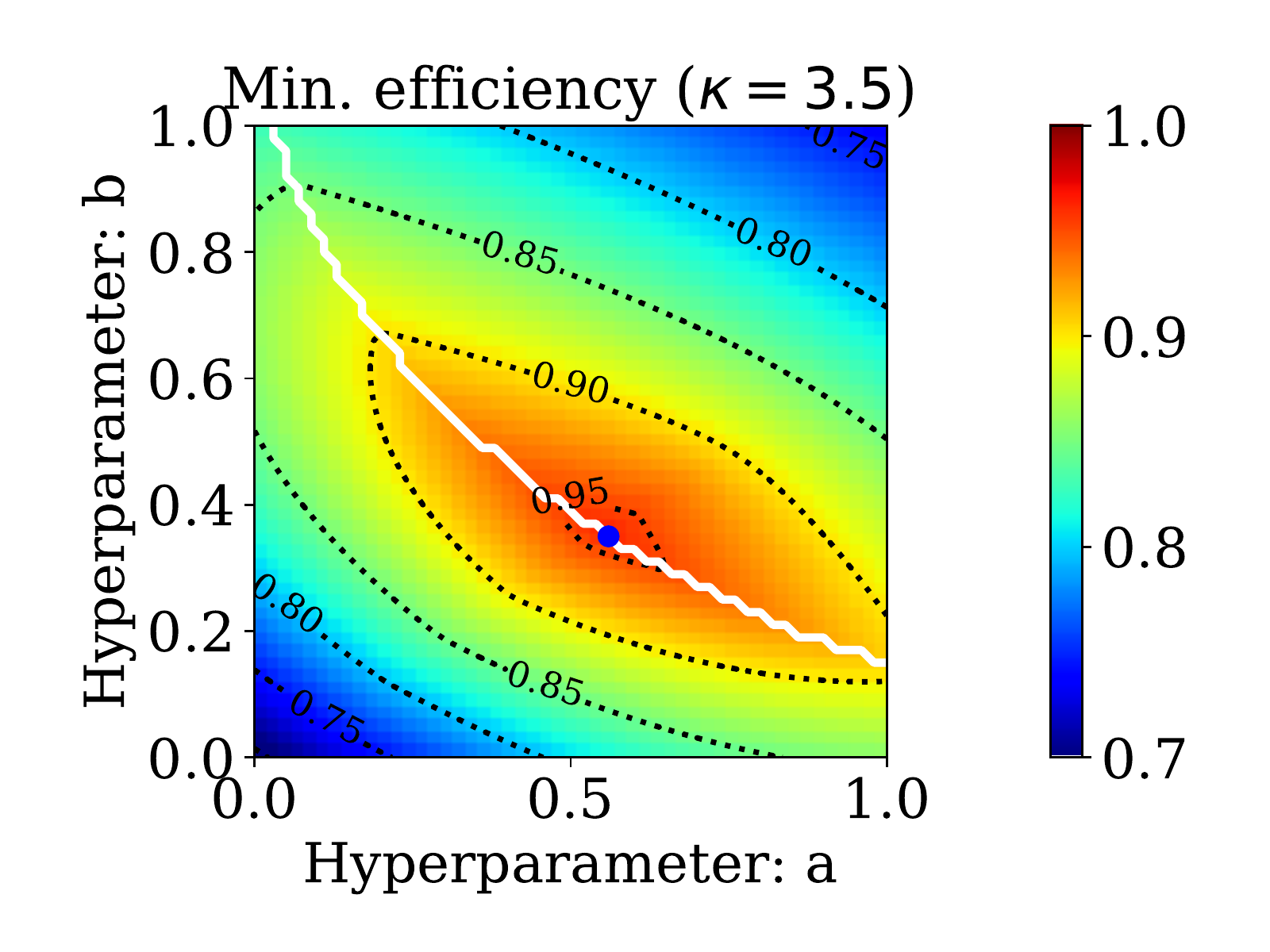}
\caption{
Minimum efficiency in $t \in [10^{-2.5},10^{2.5}]$ in the $a$-$b$ parameter space for a \MJ distribution (left) and a relativistic kappa distribution with $\kappa=3.5$ (right).
The blue circle indicates $(a,b)=(0.56, 0.35)$.
The white curve is a ridge. See the text for details.
\label{fig:hyper}}
\end{figure*}

We have searched the optimum hyperparameters in $(a, b) \in [0,1] \times [0,1]$.
For given $a$ and $b$, we calculate the total acceptance rate for every $t$ in $t \in [10^{-2.5},10^{2.5}]$ (Eqs.~\eqref{eq:MJ_eff} or \eqref{eq:rkappa_eff}). This is still a function of $t$, and then we focus on the minimum value or the lower limit of the acceptance efficiency.
As a result, we obtain the minimum efficiency for every $a$ and $b$.
We have also checked whether the rejection function always satisfies $R(x; a,b) \le 1$.
When $\max(R(x)) = R_{\rm max} > 1$, we need to normalize the rejection function, $R(x) \equiv R(x) / R_{\rm max}$. In that case, we need to rescale the total acceptance rate by $1/R_{\rm max}$, accordingly.

The two panels in Fig.~\ref{fig:hyper} display the minimum efficiency for the \MJ distribution (left) and the relativistic kappa distributions with $\kappa=3.5$ (right).
The efficiency has local maxima along the white line.
Along the white line, $R(x)$ reaches the unity, $\max(R(x)) = R_{\rm max} = 1$.
Below the white line, we renormalize the rejection function and the efficiency, as $R_{\rm max} > 1$. 
The blue circle indicates our optimum parameters, $(a,b)=(0.56,0.35)$.
The circle is virtually on the white line,
as the maximum of $R(x)$ is very close to $1$ (see the inlet in Fig.~\ref{fig:rej}).
In Fig.~\ref{fig:hyper}, the peak of the minimum efficiency is located in the south-east (north-west) vicinity of the blue circle for $\kappa \gtrsim 4.5$ ($\lesssim 4.5$),
and our circle is always in the high-efficiency region of $>95\%$.
We recommend our parameters for general purpose.

\end{document}